\documentclass[10pt, journal]{IEEEtran}

\usepackage{graphicx}             
\usepackage{caption,subcaption}            
\usepackage{url}                  
\usepackage[T1]{fontenc}
\usepackage{nicefrac}             
\usepackage{indentfirst}          
\usepackage[super, negative]{nth}
\usepackage{amsfonts}
\usepackage{amssymb}
\usepackage{multirow}
\usepackage{balance}
\usepackage{bytefield}
\usepackage{paralist}
\usepackage{cellspace}
\usepackage{algorithm}
\usepackage[noend]{algpseudocode}
\usepackage{listings}
\usepackage{subfloat}
\usepackage{etoolbox}
\usepackage[most]{tcolorbox}

\usepackage{xspace}
\usepackage{array,booktabs}
\usepackage{latexsym}
\usepackage{pifont}
\usepackage{colortbl}
\usepackage{wasysym}

\usepackage{xcolor}

\usepackage{comment}
\specialcomment{techrep}{\color{green}}{\color{black}}
\specialcomment{figtechrep}{\color{black}}{\color{black}}

\newcommand{\dfn}[1]{\textit{#1}}            
\newcommand{\revision}[1]{\textcolor{red}{#1}}
\newcommand{\ttlexceeded}{\texttt{time-exceeded}\xspace}
\newcommand{\tpropagate}{\texttt{ttl-propagate}\xspace}
\newcommand{\notpropagate}{\texttt{no-ttl-propagate}\xspace}
\newcommand{\echoreply}{\texttt{echo-reply}\xspace}
\newcommand{\echorequest}{\texttt{echo-request}\xspace}
\newcommand{\dstunreach}{\texttt{destination-unreachable}\xspace}
\newcommand{\traceroute}{\texttt{traceroute}\xspace}
\newcommand{\ping}{\texttt{ping}\xspace}

\newcommand{\tnt}{\texttt{TNT}\xspace}
\newcommand{\lt}{L$^{T}$\xspace}
\newcommand{\lter}{L$^{TE}_{R}$\xspace}
\newcommand{\lerr}{L$^{ER}_{R}$\xspace}
\newcommand{\ljter}{LJ$^{TE}_{R}$\xspace}
\newcommand{\ljerr}{LJ$^{ER}_{R}$\xspace}
\newcommand{\push}{\textsc{Push}\xspace}
\newcommand{\pop}{\textsc{Pop}\xspace}
\newcommand{\swap}{\textsc{Swap}\xspace}
\newcommand{\minttl}[2]{\textsc{Min}(#1, #2)\xspace}
\newcommand{\uturn}{\textsc{Uturn}\xspace}
\newcommand{\dupip}{\textsc{Dup\_Ip}\xspace}
\newcommand{\frpla}{\textsc{Frpla}\xspace}
\newcommand{\rtla}{\textsc{Rtla}\xspace}
\newcommand{\brpr}{\textsc{Brpr}\xspace}
\newcommand{\dpr}{\textsc{Dpr}\xspace}
\newcommand{\trtla}{$\mathcal{T}_{\rtla}$}
\newcommand{\tfrpla}{$\mathcal{T}_{\frpla}$}
\newcommand{\tuturn}{$\mathcal{T}_{\uturn}$}
\newcommand{\tlsettl}{$\mathcal{T}_{LSE\_TTL\xspace}$}
\newcommand{\unknown}{\textsc{1Hop\_Lsp}}
\newcommand{\ingnotfound}{\textsc{Ing\_Not\_Found}\xspace}
\newcommand{\tnotreached}{\textsc{Target\_Not\_Reached}\xspace}
\newcommand{\stuff}[1]{\textbf{\underline{#1}}}

\begin{document}

\definecolor{mygreen}{rgb}{0,0.6,0}
\definecolor{mygray}{rgb}{0.5,0.5,0.5}
\definecolor{mymauve}{rgb}{0.58,0,0.82}

\lstset{language=Python,
basicstyle=\scriptsize, commentstyle={\color{blue}}, frame=single,
stringstyle=\color{magenta}, numbers=left,numbersep=5pt,
numberstyle=\tiny\color{mygray}, breaklines=true, tabsize=3}

\newtcblisting{cisco}[1][mathescape=true]{size=fbox, enhanced, breakable, listing only, listing
options={style=tcblatex,basicstyle=\ttfamily\scriptsize,tabsize=2},#1}

\newtcblisting{cisco2col}[1][mathescape=true,escapeinside={(*}{*)}]{size=fbox, colback = blue!5, enhanced, breakable, listing only, listing
options={style=tcblatex,basicstyle=\ttfamily\scriptsize,tabsize=2, multicols = 2,numbers = left,  xleftmargin = 1em,
    showstringspaces = false},#1}

\title{\tnt, Watch me Explode:\\A Light in the Dark for Revealing MPLS
Tunnels}
\bstctlcite{IEEEexample:BSTcontrol}

\author{Yves Vanaubel{$^{\ast}$}, Jean-Romain Luttringer{$^{\ddag}$}, Pascal M\'erindol{$^{\ddag}$},  Jean-Jacques
Pansiot{$^{\ddag}$}, Benoit Donnet{$^{\ast}$}\\
$\ast$ Montefiore Institute, Universit\'e de Li\`ege -- Belgium\\
$\ddag$ Icube, Universit\'e de Strasbourg -- France
}



\maketitle

\begin{abstract}
Internet topology discovery has been a recurrent research topic for nearly 20
years now. Usually, it works by sending hop-limited probes (i.e., \traceroute)
towards a set of destinations to collect topological data in order to infer the
Internet topology at a given scale (e.g., at the router or the AS level).
However, \traceroute comes with multiple limitations, in particular with layer-2
clouds such as MPLS that might hide their content to \traceroute exploration.
Thus, the resulting Internet topology data and models are incomplete and
inaccurate.

In this paper, we introduce \tnt (\underline{\textbf{T}}race the
\underline{\textbf{N}}aughty \underline{\textbf{T}}unnels), an extension to
Paris traceroute for revealing most (if not all) MPLS tunnels along a path. \tnt
works in two basic stages. First, along with \traceroute probes, it looks for
evidences of the potential presence of hidden tunnels. Those evidences are
surprising patterns in the \traceroute output, e.g., abrupt and significant TTL
shifts. Second, if alarms are triggered due to the presence of such evidences,
\tnt launches additional and dedicated probing for possibly revealing the
content of the hidden tunnel. We validate \tnt through emulation with GNS3 and
tune its parameters through a dedicated measurement campaign. We also largely
deploy \tnt on the Archipelago platform and provide a quantification of tunnels,
updating so the state of the art vision of MPLS tunnels. Finally, \tnt and its
validation platform are fully and publicly available, as well as the collected
data and scripts used for processing data.
\end{abstract}

\section{Introduction}\label{intro}
For now twenty years, the Internet topology discovery has attracted a lot of
attention from the research community~\cite{survey,survey2}.  First, numerous
tools have been proposed to better capture the Internet at the IP interface
level (mainly based on \traceroute) and at the router level (by aggregating IP
interfaces of a router through \dfn{alias resolution}).  Second, the data
collected has been used to model the Internet~\cite{vespignani}, but also to
have a better knowledge of the network ecosystem and how it is organized by
operators.

However, despite the work done so far, a lot of issues still need to be fixed,
especially in data collection processes based on \traceroute.  For instance,
collecting data about Layer-2 devices connecting routers is still an open
question, although it has been addressed previously with a, nowadays, deprecated
tool (i.e., IGMP-based probing)~\cite{merlin-degree}.  Another example is the
relationship between traditional network hardware and the so-called
middleboxes~\cite{tracebox,mb_prevalence}.  Finally, MPLS
tunnels~\cite{rfc3031} also have an impact on topology discovery as they
allow to hide internal hops~\cite{mpls-ccr, mpls-invisible}.

This paper focuses on the interaction between \traceroute and MPLS.  In a
nutshell, MPLS has been designed to reduce the time required to make forwarding
decisions thanks to the insertion of \dfn{labels} (called \dfn{Label Stack
Entries}, or LSE) before the IP header.\footnote{Although MPLS can also  be used
with IPv6, we consider only IPv4 in this paper.} In an MPLS network, packets are
forwarded using an exact match lookup of a 20-bit value found in the LSE. At
each MPLS hop, the label of the incoming packet is replaced by a corresponding
outgoing label found in an MPLS switching table. The MPLS forwarding engine is
lighter than the IP forwarding engine because finding an exact match for a label
is simpler than finding the longest matching prefix for an IP address. Some MPLS
tunnels may be revealed to \traceroute because MPLS routers are able to generate
ICMP \ttlexceeded messages when the MPLS TTL expires and the ICMP message embeds
the LSE, revealing so the presence of the tunnel~\cite{mpls-sommers,mpls-ccr}.
However, MPLS supports optional features that make tunnels more or less
invisible to \traceroute. Such features modify the way routers process the IP
and MPLS TTL of a packet. By carefully analyzing some MPLS related patterns like
TTLs (e.g., the quoted forward TTL, the returned TTL of both error and standard
replies), one can identify and possibly discover the L3-hops hidden within an
MPLS cloud. A first attempt has been already proposed for revealing so-called
Invisible tunnels~\cite{mpls-invisible}. Here we are going several steps further
by providing new revelation techniques (in particular for dealing with the
ultimate hop popping feature), and its validation with multiple MPLS and BGP
configurations (by emulating MPLS network through GNS3).

This paper aims at plugging the gaps in identifying and revealing the content of
MPLS tunnels. This is done by introducing \tnt (\underline{\textbf{T}}race the
\underline{\textbf{N}}aughty \underline{\textbf{T}}unnels), an open-source
extension for Paris traceroute~\cite{parisTraceroute} including techniques for
inferring and revealing MPLS tunnels content.  Compared to our previous
work~\cite{mpls-ccr,mpls-invisible}, this paper provides multiple contributions:
\begin{enumerate}
  \item we strongly \textbf{revise the MPLS tunnel classification} as proposed
  by Donnet et al.~\cite{mpls-ccr}. In particular, when possible, we subdivide
  the ``Invisible Tunnel'' class in two more accurate categories, ``Invisible
  PHP'' and ``Invisible UHP''. We show that actually those tunnels can be
  systematically revealed when they are built with basic P2P LDP~\cite{rfc5036} or
  RSVP-TE~\cite{rfc3209} circuits (and can be at least detected if constructed
  with more complex technologies such as P2MP VPRN~\cite{rfc2764}).  We also explain
  why most ``Opaque'' tunnels content cannot be revealed in practice.  Finally,
  we refine and even correct  previous quantification of each tunnel class with
  large-scale measurements performed in the wild;
  \item we complement the state of the art with \traceroute-based
  \textbf{measurement techniques} able to reveal most (or at least detect all)
  MPLS tunnels, even those that were built for hiding their content.  While our
  previous work~\cite{mpls-invisible} required to target suspect and
  pre-analyzed zones in the Internet (i.e., considering high degree nodes and
  their neighbors visible in the ITDK dataset \cite{itdk}), we provide here
  measurement techniques fully integrated in \traceroute. We associate with each
  category of the classification \dfn{indicators} or \dfn{triggers} that are
  used to determine, on the fly, the potential presence of a tunnel and possibly
  its nature. In particular, in this paper, we are able to identify the presence
  of the newly introduced ``UHP Invisible'' tunnel class thanks to the
  duplication of an IP address in the \traceroute output.  When a trigger is
  pulled during a \traceroute exploration, an MPLS
  \dfn{revelation}~\cite{mpls-invisible} is launched with the objective of
  revealing the tunnel content.  We validate the indicators, triggers, and
  revelations using GNS-3, an emulator running the actual IOS of real routers in
  a virtualized environment\footnote{\label{footnote.gns3}See
  \url{https://gns3.com/} Note that it is also possible to emulate other router
  brands, e.g., Juniper, with GNS-3.}, on a large set of realistic
  configurations. We also show, through measurements, that our techniques are
  efficient in terms of cost (i.e., the additional amount of probes injected is
  reasonable, specially compared to the quality of new data discovered) and
  errors (false positives and false negatives);
  \item we \textbf{implement} those techniques within
  Scamper~\cite{scamper}, the state of the art network measurements
  toolbox as a Paris traceroute extension, called \tnt, and deploy it on the
  Archipelago infrastructure~\cite{ark}.  \tnt aims at replacing the old version
  of Scamper and is, thus, subject to run every day towards millions of
  destinations.  As such, we believe \tnt will be useful to study MPLS
  deployment and usage over time, increasing so our knowledge and culture on
  this technology;
  \item we \textbf{analyze} the data collected, the efficiency of \tnt in doing
  so (for tuning it to its best set of calibration parameters) and report a new
  quantification on MPLS deployment in the wild, correcting and updating so
  previous results that erroneously underestimated or overestimated
  the prevalence of some tunnel classes~\cite{mpls-ccr};
  \item we work in a \textbf{reproducibility} perspective.  As such, all our
  code (\tnt, GNS-3, data processing and analysis) as well as our collected
  dataset are made available.\footnote{\label{footnote.web}See
  \url{http://www.montefiore.ulg.ac.be/~bdonnet/mpls}}
\end{enumerate}

The remainder of this paper is organized as follows: Sec.~\ref{background}
provides the required technical background for this paper;
Sec.~\ref{taxo} revises the MPLS taxonomy initially introduced by
Donnet et al.~\cite{mpls-ccr} in the light of newly understood MPLS behaviors;
Sec.~\ref{tnt} formally introduces \tnt, our extension to \traceroute for
revealing the content of all MPLS tunnels; Sec.~\ref{tnt_calib} discusses \tnt
parameters and its calibration, while Sec.~\ref{tnt_quantif} presents results of
the \tnt deployment over the Archipelago architecture; Sec.~\ref{related}
positions our work with respect to the state of the art; finally, Sec.~\ref{ccl}
concludes this paper by summarizing its main achievements.


\section{Background}\label{background}
This section discusses the technical background required for the paper.
Sec.~\ref{background.fingerprinting} explains how hardware brand can be inferred
from collected TTLs. Sec.~\ref{background.cp} provides the basics
of MPLS labels and introduces the MPLS control plane while
Sec.~\ref{background.dp} focuses on the MPLS data plane and MPLS TTL processing.

\subsection{Network Fingerprinting}\label{background.fingerprinting}
\begin{table}[!t]
  \begin{center}
    \begin{tabular}{l|l}
      \textbf{Router Signature} & \textbf{Router Brand and OS} \\
      \hline
       $< 255, 255 >$ & Cisco (IOS, IOS XR)\\
       \hline
       $< 255, 64 >$  & Juniper (Junos)\\
       $< 128, 128 >$ & Juniper (JunosE)\\
       \hline
       $< 64, 64 >$ & Brocade, Alcatel, Linux\\
    \end{tabular}
  \end{center}
  \caption{Summary of main router signature, the first initial TTL of the pair
  corresponds to ICMP \ttlexceeded, while the second is for ICMP \echoreply.}
  \label{background.fingerprinting.table}
\end{table}

Vanaubel et al.~\cite{fingerprinting} have presented a router fingerprinting
technique that classifies networking devices based on their hardware and
operating system (OS). This method infers initial TTL values used by a router
when forging different kinds of packets. It then builds the router
\dfn{signature}, i.e., the $n$-tuple of $n$ initial TTLs. A basic pair-signature
(with $n$ = 2) simply uses the initial TTL of two different messages: an ICMP
\ttlexceeded message elicited by a \traceroute probe, and an ICMP \echoreply
message obtained from an \echorequest probe.
Table~\ref{background.fingerprinting.table} summarizes the main router
signatures, with associated router brands and router OSes. This feature is
really interesting since the two most deployed router brands, Cisco and Juniper,
have distinct MPLS behaviors and signatures.

\begin{figure}[!t]
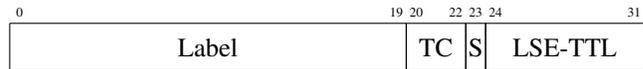

  \begin{center}
    \begin{bytefield}[bitwidth=0.75em]{32}
      \bitheader{0,19,20,22,23,24,31}\\
      \bitbox{20}{Label} \bitbox{3}{TC}
      \bitbox{1}{S} \bitbox{8}{LSE-TTL}
    \end{bytefield}
  \end{center}
  \caption{The MPLS label stack entry (LSE) format.}
  \label{background.cp.lse}
\end{figure}

\subsection{MPLS Basics and Control Plane}\label{background.cp}
MPLS routers, i.e., \dfn{Label Switching Routers} (LSRs), exchange labeled
packets over \dfn{Label Switched Paths} (LSPs). In practice, those packets are
tagged with one or more \dfn{label stack entries} (LSE) inserted between the
frame header (data-link layer) and the IP packet (network layer). Each LSE is
made of four fields as illustrated by Fig.~\ref{background.cp.lse}:
an MPLS label used for forwarding the packet to the next router, a Traffic Class
field for quality of service, priority, and Explicit Congestion
Notification~\cite{rfc5462}, a bottom of stack flag bit (to indicate whether the
current LSE is the last in the stack~\cite{rfc3032})\footnote{To simplify the
presentation we will consider  only one LSE in the remainder of this paper}, and
a time-to-live (LSE-TTL) field having the same purpose as the IP-TTL
field~\cite{rfc3443} (i.e., avoiding routing loops).

Labels may be allocated through the \dfn{Label Distribution Protocol}
(LDP)~\cite{rfc5036}. Each LSR announces to its neighbors the association
between a prefix in its routing table and a label it has chosen for a given
\dfn{Forwarding Equivalent Class} (a FEC is a destination prefix by default),
populating so a \dfn{Label Forwarding Information Table} (LFIB) in each LSR.
With LDP, a router advertises the same label to all its neighbors for a given
FEC.
LDP is mainly used for scalability reasons (e.g., to limit BGP-IGP interactions
to edge routers) and to avoid anomalies for the transit traffic such as iBGP
deflection issues. Indeed, LDP deployed tunnels use the routes computed by
the IGP (without any interest at the first, and naive, glance) as the LFIB is
built on top of the IGP FIB. Labels can also be distributed through
RSVP-TE~\cite{rfc3209}, when MPLS is used for Traffic Engineering (TE) purposes.
In practice, most operators deploying RSVP-TE tunnels use
LDP~\cite{mpls-invisible} as a default labeling protocol.

With LDP, MPLS has two ways of binding labels to destination prefixes:
\begin{enumerate}
  \item through ordered LSP control (default configuration of Juniper
  routers~\cite{junos});
  \item through independent LSP control (default configuration of Cisco
  routers~\cite[Chap. 4]{cisco_lsettl}).
\end{enumerate}
In the former mode, a LSR only binds a label to a prefix if this prefix is local
(typically, the exit point of the LSR), or if it has received a label binding
proposal from the IGP next hop towards this prefix. This mode is thus iterative
as each intermediate upstream LSR waits for a proposal of its downstream LSR (to
build the LSP from the exit to the entry point).  Juniper routers use this mode
as default and only propose labels for loopback IP addresses. In the second
mode, the Cisco default one, a LSR creates a label binding for each prefix it
has in its RIB (connected or -- redistributed in -- IGP routes only) and
distributes it to all its neighbors. This mode does not require any proposal
from downstream LSRs.  Consequently, a label proposal is sent to all neighbors
without ensuring that the LSP is enabled up to the exit point of the tunnel. LSP
setup takes less time but may lead to uncommon situations in which an LSP can end
abruptly before reaching the exit point (see Sec.~\ref{taxo} for details.)

The last LSR towards a FEC is the \dfn{Egress Label Edge Router} (the Egress LER
-- PE$_2$ in Fig.~\ref{background.dp.fig}). Depending on its configuration, two
labeling modes may be performed.  The default mode~\cite{mpls-invisible} is
\dfn{Penultimate Hop Popping} (PHP), where the Egress advertises an Implicit
NULL label (label value of 3~\cite{rfc3032} -- see
Table~\ref{background.dp.swap.reserved} for details on reserved MPLS label
values). The previous LSR (\dfn{Penultimate Hop LSR},PH -- P$_3$ in
Fig.~\ref{background.dp.fig}) is in charge of removing the LSE to reduce the
load on the Egress. In the \dfn{Ultimate Hop Popping} (UHP), the Egress LER
advertises an Explicit NULL label (label value of 0~\cite{rfc3032} -- see
Table~\ref{background.dp.swap.reserved} for details on reserved MPLS label
values). The PH will use this Explicit NULL label and the Egress LER will be
responsible for its removal. Labels assigned by LSRs other than the Egress LER
are distinct from Implicit or Explicit NULL labels. The \dfn{Ending Hop LSR}
(EH) is the LSR in charge of removing the label, it can be the PH in case of
PHP, the Egress LER in case of UHP or possibly another LSR in the case of
independent LSP control.

Table~\ref{background.cp.terminology} provides a summary of main acronyms used
by MPLS and their corresponding concept in the classic IP world.

\begin{table}[!t]
  \begin{center}
    \begin{tabular}{l|l|l}
      \textbf{Acronym} & \textbf{Meaning} & \textbf{IP}\\
      \hline
      LSR & \stuff{L}abel \stuff{S}witching \stuff{R}outer & \multirow{3}{*}{Router}\\
      PH LSR & \stuff{P}enultimate \stuff{H}op LSR         & \\
      EH  & \stuff{E}nding \stuff{H}op LSR                 & \\
      \hline
      LER & \stuff{L}abel \stuff{E}dge \stuff{R}outer    & Border Router\\
      LSP & \stuff{L}abel \stuff{S}witching \stuff{P}ath & Tunnel\\
      \hline
      LSE & \stuff{L}abel \stuff{S}ack \stuff{E}ntry & Header\\
      LSE-TTL & LSE \stuff{T}ime-\stuff{t}o-\stuff{L}ive & IP-TTL\\
      \hline
      LDP & \stuff{L}abel \stuff{D}istribution \stuff{P}rotocol & \multirow{2}{*}{Signaling}\\
      RSVP-TE & \stuff{R}e\stuff{S}er\stuff{V}etation \stuff{P}rotocol -- \stuff{T}raffic \stuff{E}ngineering & \scriptsize{(control plane)}\\
      \hline
      LIB  & \stuff{L}abel \stuff{I}nformation \stuff{B}ase                    & RIB\\
      LFIB & \stuff{L}abel \stuff{F}orwarding \stuff{I}nformation \stuff{B}ase & FIB\\
      \hline
      PHP & \stuff{P}enultimate \stuff{H}op \stuff{P}opping & \multirow{2}{*}{Decapsulation}\\
      UHP & \stuff{U}ltimate \stuff{H}op \stuff{P}opping    & \\
      \hline
      FEC & \stuff{F}orwarding \stuff{E}quivalent \stuff{C}lass & QoS Class\\
    \end{tabular}
  \end{center}
  \caption{MPLS terminology with its matching in the classic IP world.}
  \label{background.cp.terminology}
\end{table}

\subsection{MPLS Data Plane and TTL processing}\label{background.dp}
\begin{figure*}[!t]
  \begin{center}
    \includegraphics[width=16cm]{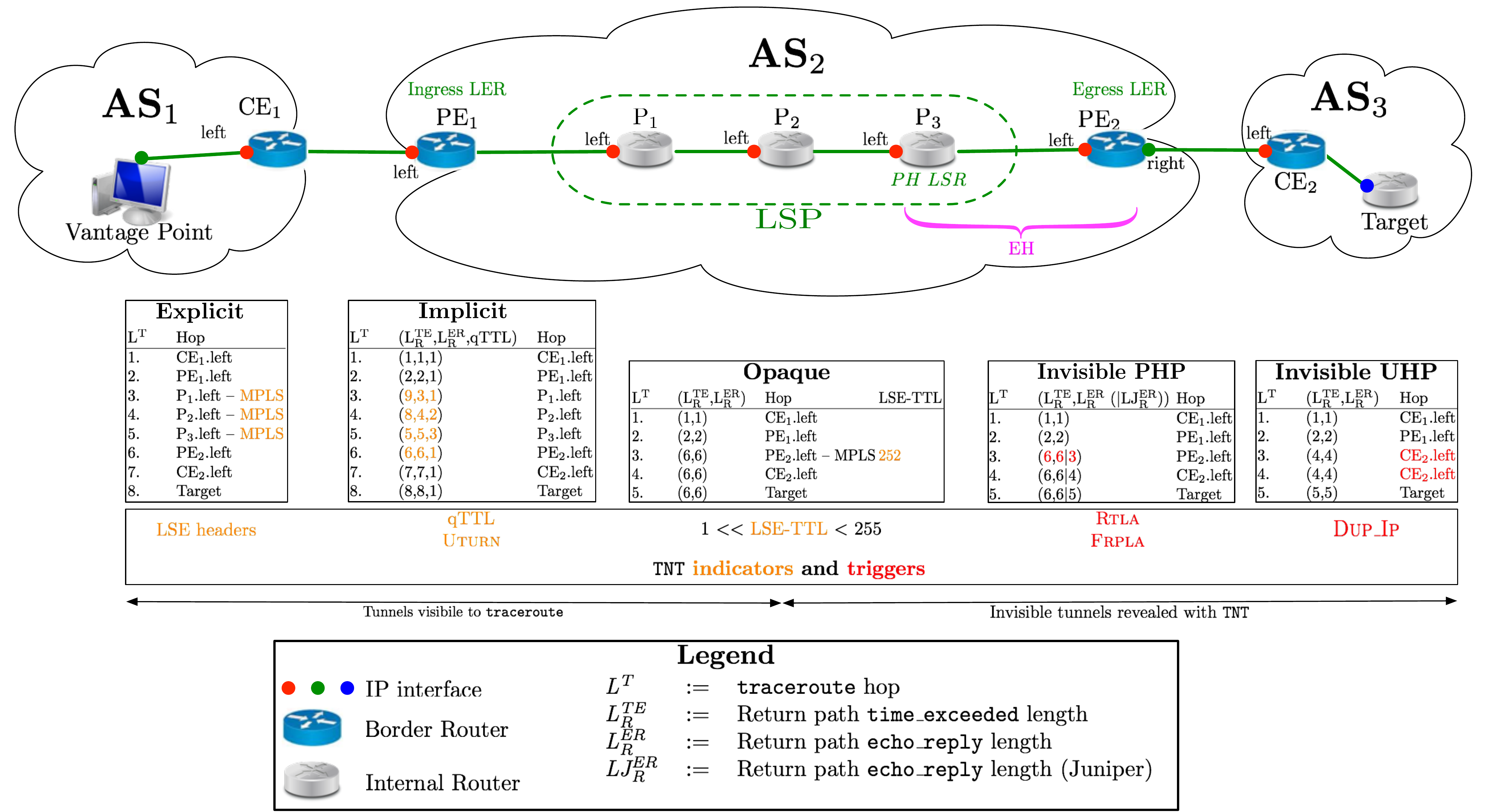}
  \end{center}
  \caption{Illustration of MPLS vocabulary and relationship between MPLS and
  \traceroute. The figure is made of three parts.  The upper part represents the
  network topology we use, throughout the paper to illustrate concepts.  In
  particular, with respect to MPLS, P$_1$ is the LSP First Hop (FH), while P$_3$ is
  the Penultimate Hop (PH). In case of PHP, P$_3$ is the Ending Hop (EH)
  and  is responsible for removing the LSE. In case of UHP, the LSE is removed
  by the Egress LER (PE$_2$). The middle part of the figure presents the MPLS
  Tunnel classification, as observed with \traceroute (this classification
  is a revision of the original one proposed by Donnet et al.) Finally, the
  bottom part of the figure provides triggers and indicators of an MPLS tunnel
  presence when probing with \tnt. The relationship between the
  trigger/indicator and the observation made with probing is provided in red.
  Additional information (such as \ttlexceeded path length) are provided. This
  is used in Sec.~\ref{tnt} for illustrating \tnt.
  }
  \label{background.dp.fig}
\end{figure*}

Depending on its location along the LSP, a LSR applies one of the three
following operations:
\begin{itemize}
  \item \push (Sec.~\ref{background.dp.push}).  The first MPLS router, i.e., the
  tunnel entry point pushes one or several LSEs in the IP packet that turns into
  an MPLS one. The \dfn{Ingress Label Edge Router} (Ingress LER) associates the
  packet FEC to its LSP.
  \item \swap (Sec.~\ref{background.dp.swap}).  Within the LSP, each LSR makes a
  label lookup in the LFIB, swaps the incoming label with its corresponding
  outgoing label, and sends the MPLS packet further along the LSP.
  \item \pop (Sec.~\ref{background.dp.pop}).  The EH, the last LSR of the LSP,
  deletes the LSE, and converts the MPLS packet back into an IP one. The EH can
  be the \dfn{Egress Label Edge Router} (the Egress LER) when UHP is enabled or
  the PH otherwise.
\end{itemize}

Fig.~\ref{background.dp.fig} (above part) illustrates the main vocabulary
associated to MPLS tunnels.

\subsubsection{LSP Entry Behavior}\label{background.dp.push}
When an IP packet enters an MPLS cloud, the Ingress LER binds a label to the
packet thanks to a lookup into its LFIB, depending on the packet FEC, e.g., its
IP destination prefix.  Prior to pushing the LSE into the packet, the Ingress
LER has to initialize the LSE-TTL (see Fig.~\ref{background.cp.lse}).
Two behaviors can then be configured: either the Ingress LER resets the LSE-TTL
to an arbitrary value (255,  \notpropagate) or it copies the current IP-TTL value
into the LSE-TTL (\tpropagate, the default behavior).  Operators can configure
this operation using the \notpropagate option provided by the router
manufacturer~\cite{rfc3443}. In the former case, the LSP is called a \dfn{pipe
LSP}, while, in the latter case, a \dfn{uniform} one.

Once the LSE-TTL has been initialized, the LSE is pushed on the packet and then
sent to an outgoing interface of the Ingress LER. In most cases, except for a
given Juniper OS (i.e., Olive), the IP-TTL is decremented before being
encapsulated into the MPLS header.

\subsubsection{LSP Internal Behavior}\label{background.dp.swap}
Upon an MPLS packet arrival, an LSR decrements its LSE-TTL.  If it does not
expire, the LSR looks up the label in its LFIB.  It then swaps the top LSE with
the one provided by the LFIB. The operation is actually a swap only if the
outgoing label returned by the LFIB is neither Implicit NULL nor
empty\footnote{In practice the actual label used for the forwarding is then
greater than or equal to 0 (this specific value being reserved for Explicit NULL
tunnel ending, i.e. for UHP) but excluding by design the reserved value $3$ that
is dedicated for Implicit NULL.}. Otherwise, it is a \pop operation as described
in the next subsection. Finally, the packet is sent to the outgoing interface of
the LSR with a new label, both according to the LFIB.

If the LSE-TTL expires, the LSR, in the fashion of any IP router, forges an ICMP
\ttlexceeded that is sent back to the packet originator.  It is worth to notice
that a LSR may implement RFC 4950~\cite{rfc4950} (as it should be the case in
all recent OSes).  If so, it means that the LSR will quote the full MPLS LSE
stack of the expired packet in the ICMP \ttlexceeded message.

ICMP processing in MPLS tunnels varies according to the ICMP type of message.
ICMP \dfn{Information messages} (e.g., \echoreply) are directly sent to the
destination (e.g., originator of the \echorequest) if the IP FIB allows for it
(otherwise no replies are generated). On the contrary, ICMP \dfn{Error messages}
(e.g., \ttlexceeded) are generally forwarded to the Egress LER that will be in
charge of forwarding the packet through its IP plane~\cite{mpls-ccr}.
Differences between Juniper and Cisco OS and configurations are discussed in
details in Sec.~\ref{tnt.validation}.

\begin{table}[!t]
  \begin{center}
    \begin{tabular}{l|l|l}
      \textbf{Value} & \textbf{Meaning} & \textbf{Mechanism}\\
      \hline
      \multirow{3}{*}{0} & \multirow{3}{*}{IPv4 Explicit NULL} & downstream LSR should pop label\\
                         &                    & immediately (UHP -- popped packet\\
        				 &                    & is an IPv4 datagram)\\
      \hline
      1 & Router Alert       & deliver to control plane, do not forward\\
      \hline
      \multirow{3}{*}{2} & \multirow{3}{*}{IPv6 Explicit NULL} & downstream LSR should pop label\\
                         &                    & immediately (UHP -- popped packet\\
        				 &                    & is an IPv6 datagram)\\
	  \hline
      \multirow{3}{*}{3} & \multirow{3}{*}{Implicit NULL} & pop immediately and treat as an IPv4\\
                         &                                & packet (PHP)\\
    \end{tabular}
  \end{center}
  \caption{Reserved label values in MPLS and their meanings.}
  \label{background.dp.swap.reserved}
\end{table}

\subsubsection{LSP Exit Behavior}\label{background.dp.pop}
At the MPLS packet arrival, the EH decrements the LSE-TTL. If this TTL
does not expire, the EH then pops the LSE stack after having determined the new
IP-TTL.

Applying PHP comes with the advantage of reducing the load on the Egress LER,
especially if it is the root of a large LSP-tree. This means that, when using
PHP, the last MPLS operation (i.e., \pop) is performed one hop before the Egress
LER, on the PH. On the contrary, UHP is generally used only when the ISP
implements more sophisticated traffic engineering operations or wants to make
the tunnel content and semantics more transparent to the customers.\footnote{The
UHP feature has been recently made available  on Juniper routers when LSPs are
set with LDP.  However, PHP remains the rule on
Juniper~\cite[Chap.~1]{juniper_php}.}

When leaving a tunnel, the router has to decide which is the correct TTL value
(IP-TTL or LSE-TTL) to copy in the IP header.  If the Ingress LER has activated
the \notpropagate option, the EH should pick the IP-TTL of the incoming packet
while the LSE-TTL should be selected otherwise. This way, the resulting outgoing
TTL cannot be greater than the incoming one: in the former case, internal hops
are not counted because the tunnel is hidden while they are for the latter case.
In both cases, the TTL behavior remains monotonic. In order to synchronize both
ends of the tunnel without any message exchange, two mechanisms might be used
for selecting the IP-TTL at the EH:
\begin{enumerate}
  \item applying a \minttl{IP-TTL}{LSE-TTL} operation (solution implemented for
  Cisco PHP configurations~\cite{cisco_lsettl});
  \item assuming  the Ingress configuration (\tpropagate or not) is the same as
  the local configuration (solution implemented by some JunOS and also in some
  Cisco UHP configuration).
\end{enumerate}
Applying the \minttl{IP-TTL}{LSE-TTL} is the best option because it correctly
supports heterogeneous \tpropagate configurations in any case while, at the same
time, mitigating forwarding loop without exchanging signalization messages.

This \minttl{IP-TTL}{LSE-TTL} behavior might be used for detecting the presence
of hidden MPLS tunnels \cite{mpls-invisible}.  Indeed, it is likely that the EH
generating the ICMP \ttlexceeded message will use the same MPLS cloud back to
reply to the vantage point.  In that case, when the reply will leave the MPLS
cloud, the returning EH ($P_1$ in Fig.~\ref{background.dp.fig}) will choose to
copy the LSE-TTL in the IP-TTL, as the IP-TTL has been initialized at its
maximum value on the Egress of the forward tunnel ($255$ for a Cisco router --
see Sec.~\ref{background.fingerprinting}).  As a consequence, while the forward
path hides the MPLS cloud because the \minttl{IP-TTL}{LSE-TTL} operated on the
forward PH ($P_3$) selects the IP-TTL which is lower, the return path indicates
its presence because the returning PH ($P_1$), on the contrary, selects the
LSE-TTL.  In general, a sufficient condition for this pattern to occur is if the
returning Ingress, which is the forward EH, re-uses the MPLS cloud back.

In practice, it is interesting to mention that this MPLS behavior is strongly
dependent on the implementation and the configuration.  For instance, on some
Juniper OS routers (at least with JunOS Olive) or when the UHP option is
activated on some Cisco IOS (at least with the 15.2 version), the
\minttl{IP-TTL}{LSE-TTL} operation is not -- systematically -- applied. The EH
assumes that the propagation configuration is homogeneous among LERs. When it is
not the case (\tpropagate at one end of the tunnel and \notpropagate at the
other end), the PH (for PHP routers without \minttl{IP-TTL}{LSE-TTL}) or the
Egress LER (for the Cisco UHP configuration) will use the IP-TTL instead of the
LSE-TTL, leading so to a so-called \dfn{jump} effect with \traceroute (i.e., as
many hops as the LSP length are skipped after the tunnel).  Except when
explicitly stated, we will consider homogeneous configurations (e.g.,
\tpropagate on the whole tunnel) in the remainder of the paper. Finally, it is
worth noticing that mixing UHP and PHP (hybrid configurations) can also result
in uncommon behaviors.\footnote{Those behaviors are described and
discussed in details in the Appendix, at the end of this paper.}


\section{Revisiting MPLS Tunnels Taxonomy}\label{taxo}
According to whether LSRs implement RFC4950 or not
(Sec.~\ref{background.dp.swap}) and whether they activate the \tpropagate option
or not (Sec.~\ref{background.dp.push}), MPLS tunnels can be revealed to
\traceroute following Donnet et al.~\cite{mpls-ccr} taxonomy.

\dfn{Explicit} tunnels are those with RFC4950 and the \tpropagate option
activated (this is the default configuration). As such, they are fully visible
with \traceroute, including labels along the LSP.  \dfn{Implicit} tunnels
activate the \tpropagate option but do not implement the RFC4950. No IP
information is missed but LSRs are viewed as ordinary IP routers, leading to a
lack of ``semantic'' in the \traceroute output. \dfn{Opaque} tunnels are
obscured from \traceroute as the \tpropagate option is disabled while the
RFC4950 is implemented and, more decisive, the EH that pops the last label has
not received an Explicit or Implicit NULL proposal for the given FEC (making the
LSP end in a non controlled fashion). Consequently, the EH can be seen while the
remainder of the tunnel is hidden.  Finally, \dfn{Invisible} tunnels are hidden
as the \notpropagate option is activated (RFC4950 may be implemented or not).

As illustrated in Fig.~\ref{background.dp.fig} (middle part), Explicit tunnels
are the ideal case as all the MPLS information comes natively with \traceroute.
For Implicit tunnels, Donnet et al.~\cite{mpls-ccr} have proposed techniques for
identifying the tunnel based on the way LSRs process ICMP messages (see
Sec.~\ref{background.dp.swap} -- the so-called \uturn) and the IP-TTL quoted in
the \ttlexceeded message (the so-called qTTL) that is increased by one at each
subsequent LSR of the LSP due to the \tpropagate option (ICMP \ttlexceeded are
generated based on the LSE-TTL while the IP-TTL of the probe is left unchanged
within the LSP and, thus, quoted as such in the ICMP \ttlexceeded).

Opaque tunnels are only encountered with Cisco LSPs and are a consequence of the
way labels are distributed with LDP (see Sec.~\ref{background.cp}). Indeed, a
label proposal may be sent to all neighbors without ensuring that the LSP is
enabled up to the Egress LER, leading so to Opaque tunnels because an LSP can
end abruptly without reaching the Egress LER (where the prefix is injected in
the IGP) that should bind an Explicit (UHP) or Implicit NULL label (PHP).  As
illustrated in Fig.~\ref{background.dp.fig}, Opaque tunnels and their length can
be identified thanks to the LSE-TTL. These LSPs end without a standard terminating
label (Implicit or Explicit NULL) and so they \textit{break} with the last MPLS
header of the neighbor that may not be an MPLS speaker. Thanks to our large
scale campaign and cross-validation with our emulation platform, we realized
that the vast majority of Opaque tunnels seems to be caused by
Carrier-of-Carriers VPN~\cite{rfc4364} or similar technologies. Indeed, they
provoke an abrupt tunnel ending (the bottom label is necessarily carried up to
the end of the tunnel to determine the correct outgoing VPN), and unfortunately
lead to non revealable tunnels as we will show later.

The \traceroute behavior, for Invisible tunnel, is different according to the
way the LSE is popped from the packet (i.e., UHP or PHP), as illustrated in
Fig.~\ref{background.dp.fig}.   Invisible tunnels are problematic, as they lead
to a false vision of the Internet topology, creating false links, and spoiling
graph metrics, such as the node degree distribution~\cite{mpls-invisible}. In
this paper, we revisit the original taxonomy by doing a clear distinction
between Invisible tunnels produced with PHP and UHP. In Donnet et
al.~\cite{mpls-ccr},  the class ``Invisible'' only covered PHP.
Vanaubel et al.~\cite{mpls-invisible} have since proposed techniques for
revealing the content of Invisible MPLS tunnels only in the case of PHP.

With Invisible UHP tunnels, the behavior is clearly different, at least for
Cisco routers using the 15.2 IOS\footnote{While it is now possible to enable UHP
with Juniper for LDP, \tnt is not able to make a distinction between the two
because the visible and revealed patterns are the same.}. Upon reception of a
packet with IP-TTL of 1, the Egress LER does not decrement this TTL, but,
rather, forwards the packet to the next hop ($CE_2$ in the example), so that the
Egress does not show up in the trace. In contrast, the next hop will appear
twice: once for the probe that should have expired at the Egress and once at the
next probe. UHP indeed provokes a surprising pattern, a duplicated IP at two
successive hops, illustrated as ``Invisible UHP'' in
Fig.~\ref{background.dp.fig}. This duplicated IP addresses might be
misunderstood as a forwarding loop.

On the contrary, PHP moves the \pop function at the PH, one hop before the
tunnel end. This PH does not decrement the IP-TTL whatever its value is.
Except for some JunOS, the packet is still MPLS switched because the LSE-TTL has
not expired on it. It is somehow surprising because for Explicit and Implicit
tunnels, the PH replies on its own. It is because, in this cases, the LSE-TTL has also expired.
In Fig.~\ref{background.dp.fig}, we can see that there is no more asymmetry in
path length for router P$_3$ proving so its reply does not follow a \uturn via
the Egress. On the contrary, any other LSR on the LSP builds a \ttlexceeded
message when the LSE-TTL expires and then continues to MPLS switch their reply
error packet to the Egress LER unless the \texttt{mpls ip ttl-expiration pop
<stack size>} command has been activated for Cisco routers. It seems to be just
an option for Juniper routers with the \texttt{icmp-tunneling} command.

Note that Opaque and Invisible UHP tunnels are Cisco tunnels (signature
$<255,255>$) due to specific implementations. Invisible PHP are Juniper
(signature $<255,64>$), Linux boxes (signature $<64,64>$), or Cisco tunnels but
they do not behave exactly the same as we will explain later.

Sec.~\ref{tnt} extends techniques for revealing MPLS tunnels by proposing
and implementing integrated measurement techniques for all tunnels (i.e.,
Explicit, Implicit, Opaque, and both UHP and PHP Invisible ones) in a single
tool called \tnt.


\section{\tnt Design and Reproducibility}\label{tnt}
This section introduces our tool, \tnt (\underline{\textbf{T}}race the
\underline{\textbf{N}}aughty \underline{\textbf{T}}unnels), able to reveal MPLS
tunnels along a path.  \tnt is an extension to Paris
Traceroute~\cite{parisTraceroute} so that we avoid most of the problems related
to load balancing.  \tnt has been implemented within Scamper~\cite{scamper}, the
state-of-the-art network measurements toolbox, and is freely
available.\textsuperscript{\ref{footnote.web}} Sec.~\ref{tnt.overview} provides
an overview of \tnt, while Sec.~\ref{tnt.trig} and Sec.~\ref{tnt.revelation}
focus on techniques for revealing hidden tunnels and how those techniques are
triggered. Finally, Sec.~\ref{tnt.validation} explains how we validated \tnt on
a GNS-3 platform\textsuperscript{\ref{footnote.gns3}}, an emulator running the
actual OS of real routers in a virtualized environment.

\subsection{Overview}\label{tnt.overview}
\begin{lstlisting}[escapeinside={(*}{*)}, label=tnt.overview.trace_tunnel,
caption=Pseudo-code for \tnt]
Codes := 0, None ; 1, LSE ; 2, qTTL ; 3, (*\uturn*); 4, LSE-TTL;
 5, (*\frpla*); 6, (*\rtla*); 7, (*\dupip*).
trace_naughty_tunnel(target):
	prev_hop, cur_hop, next_hop = None

	for (ttl=STARTING_TTL, !halt(ttl, target), ttl++)
		state, tun_code = None
		next_hop = trace_hop(ttl)

		#first check uniform tunnel evidence with indicators
		tun_code = check_indicators(cur_hop) (*\label{check_indicators}*)
		#possibly fires (*\tnt*) with triggers or opaques tunnels
		if (tun_code == None)
			tun_code = check_triggers(prev_hop, cur_hop, next_hop)
			#check if cur_hop does not belong to a uniform LSP
			if (tun_code != None)
				#potential hidden tunnel to reveal
				state = reveal_tunnel(prev_hop, cur_hop, tun_code)
		elif (tun_code == LSE-TTL)
			#potential opaque tunnel to reveal
			state = reveal_tunnel(prev_hop, cur_hop, tun_code)

		#hop by hop and tunnel display
		dump(cur_hop, tun_code, state)

		#sliding pair of IP addresses
		prev_hop = cur_hop #candidate ingress LER
		cur_hop = next_hop #candidate egress LER
\end{lstlisting}

\tnt is conceptually illustrated in Listing~\ref{tnt.overview.trace_tunnel}. At
the macroscopic scale, the \texttt{trace\_naughty\_tunnel()} function is a
simple loop that fires probes towards each processed target. \tnt consists in
collecting, in a hop-by-hop fashion,
intermediate IP addresses (\texttt{trace\_hop()} function) between the vantage
point and the target. Tracing a particular destination ends when the
\texttt{halt()} function returns true: the target has been reached or a gap has
been encountered (e.g., five consecutive non-responding hops). \tnt uses a
moving window of two hops such that, at each iteration, it considers a potential
Ingress LER (i.e., \texttt{prev\_hop}) and a potential Egress LER (i.e.,
\texttt{cur\_hop}) for possibly revealing an Invisible tunnel between them.
Indicators allow to check if the current hop does not belong to a uniform
tunnel, i.e., a visible one (see line~\ref{check_indicators}).

For each pair of collected IP addresses with \texttt{trace\_hop()}, \tnt checks
for the presence of tunnels through so called \dfn{indicators} and
\dfn{triggers}. The former provides reliable indications about the presence of
an MPLS tunnel without necessarily requiring additional probing. Generally,
indicators correspond to uniform tunnels (or to the last hop of an Opaque
tunnel), and are, mostly, basic evidence of visible MPLS presence such, as LSEs
quoted in the ICMP \ttlexceeded packet (see Sec.~\ref{tnt.trig} for details).
Triggers are mainly unsigned values suggesting the potential presence of
Invisible tunnels through a large shifting in path length asymmetry (see
Sec.~\ref{tnt.trig} for details). When exceeding a given threshold
$\mathcal{T}$, such triggers fire path revelation methods (function
\texttt{reveal\_tunnel()}) between the potential Ingress and Egress LERs as
developed in Sec.~\ref{tnt.revelation}. If intermediate hops are found, they are
stored in a global stack structure named \texttt{revealed\_lsrs}.

\texttt{STARTING\_TTL} is  a parameter used to avoid tracing repeatedly the
nodes close to the vantage point~\cite{dt}, usually \texttt{STARTING\_TTL} $\in
[3,5]$.

Finally, at each loop iteration, the collected data is dumped into a warts file,
the Scamper file format for storing IPv4/IPv6 \traceroute records. This job is
performed by the \texttt{dump()} function. It writes potential revealed hops
(available in the global stack structure \texttt{revealed\_lsrs}), and any
useful information, such as tags, identifying the tunnel's type\revision{,} and
revelation method, if any.

\subsection{Indicators and Triggers}\label{tnt.trig}
\begin{lstlisting}[escapeinside={(*}{*)}, label=tnt.trig.check_indicators,
caption=Pseudo-code for checking indicators]
code check_indicators(hop):
	#hop must exist
	if (hop == None)
		return None

	if (is_mpls(hop))
    if ((*\tlsettl*) < hop.lse_ttl < 255) (*\label{lse_opaque}*)
      #opaque tunnel are both indicators and triggers
      return LSE-TTL (*\label{lse_opaque_end}*)
    else
      #explicit tunnel
		  return LSE(*\label{explicit_lse}*)

	if (hop.qttl > 1)(*\label{qttl}*)
		#implicit tunnel
		return qTTL(*\label{exception2}*)

	#retrieve path length from raw TTLs
	(*\lter*) = path_len(hop.ttl_te)
	(*\lerr*) = path_len(hop.ttl_er)

	#(*\uturn*) will be turned into (*\rtla*) for junOS signatures
	if (|(*\lter*) - (*\lerr*)| (*$>$ \tuturn*) && !signature_is_junOS(hop))(*\label{exception}*)
	#implicit tunnel
		return (*\uturn*)

	return None
\end{lstlisting}

Tunnels indicators are pieces of evidence of MPLS tunnel presence and concern
cases where tunnels (or parts of them) can be directly retrieved from the
original \traceroute. They are used for Explicit tunnels and uniform/visible
tunnels in general.  Explicit tunnels are indicated through LSEs directly quoted
in the ICMP \ttlexceeded message -- See line~\ref{explicit_lse} in
Listing~\ref{tnt.trig.check_indicators} and \traceroute output on
Fig.~\ref{background.dp.fig}.  It is worth noting that
Fig.~\ref{background.dp.fig} highlights the main patterns \tnt looks for firing
or not additional path revelation in a simple scenario where forward and return
paths are symmetrical.

The indicator for Opaque tunnels consists in a single hop LSP with the quoted
LSE-TTL not being equal to $1$, due to the way labels are distributed within
Cisco routers (see Sec.~\ref{background.cp}) or the way Cisco routers deal with
VPRN tunnel ending.\footnote{Juniper routers never lead to Opaque indicators
because they behave differently as discussed in Sec.~\ref{tnt.validation}}  This
is illustrated in Fig.~\ref{background.dp.fig} where we get a value of $252$
because the LSP is actually $3$ hops long. This surprising quoted LSE-TTL is an
evidence in itself. It is illustrated in lines~\ref{lse_opaque}
to~\ref{lse_opaque_end} in Listing~\ref{tnt.trig.check_indicators}, where a hop
is tagged as Opaque if the quoted LSE-TTL is between a minimum threshold,
\tlsettl (see Sec.~\ref{tnt_calib} for fixing a value for the threshold) and 254
(LSE-TTL is initialized to 255~\cite{rfc3443}). Note that this pattern resulting
from an Opaque tunnel is both an indicator and a trigger: \tnt passively
understands the tunnel is incomplete and try to reveal its content with new
active measurements.

Implicit tunnels are detected through qTTL and/or \uturn
indicators~\cite{mpls-ccr}. First, if the IP-TTL quoted in an ICMP \ttlexceeded
message (qTTL) is greater than one, it likely reveals the \tpropagate option at
the Ingress LER of an LSP. For each subsequent \traceroute probe within the LSP,
the qTTL will be one greater, resulting in an increasing sequence of qTTL
values. This indicator is considered in line~\ref{qttl} in
Listing~\ref{tnt.trig.check_indicators}. Second, the \uturn indicator relies on
the fact that, by default, LSRs send ICMP \ttlexceeded messages to the Egress
LER which, in turns, forwards the packets to the probing source.
However, they reply directly to other kinds of probes (e.g., \echorequest) using
their own IP forwarding table, if available. As a result, return
paths are generally shorter for \echoreply messages than for \ttlexceeded replies.
Thereby, \uturn is the signature related to the difference in these lengths.
This is illustrated in Fig.~\ref{background.dp.fig} (Implicit and Explicit
tunnels follow the same behavior except for  RFC4950 implementation).
On P$_1$, we have \uturn(P$_1$) = \lter - \lerr = 9 - 3 = 6. With a symmetric
example, one can formalize the \uturn pattern for an LSR P$_i$ in an LSP of
length $LL$ as follows:
\begin{equation}
\uturn(P_i) = 2 \times (LL - i + 1) .
\label{eqn.uturn}
\end{equation}

\begin{figure*}[!t]
  \begin{center}
    \begin{subfigure}[b]{8.2cm}
      \includegraphics[width=8.2cm]{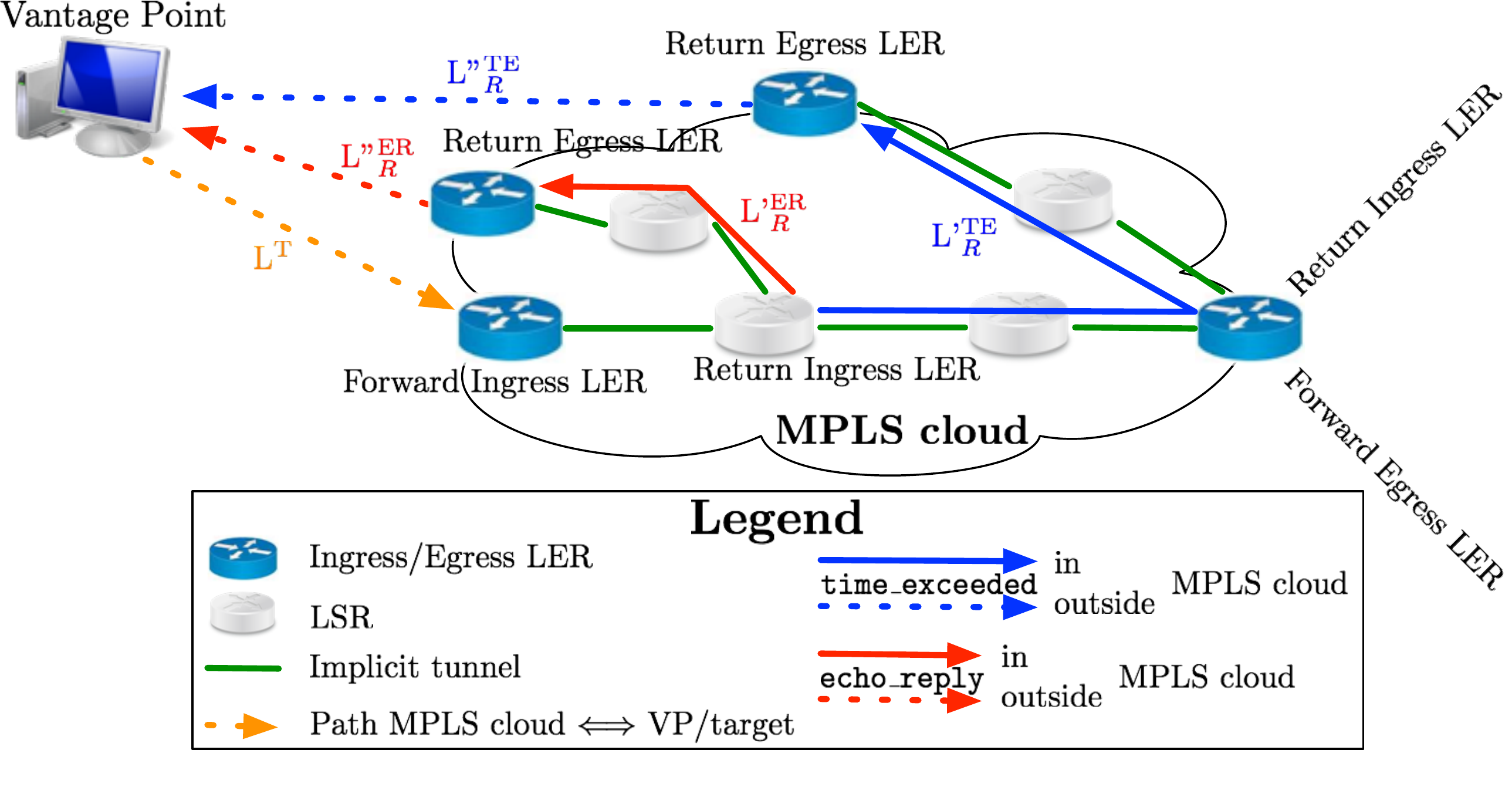}
      \caption{Implicit tunnels.}
      \label{tnt.trig.asym.implicit}
    \end{subfigure}
    \begin{subfigure}[b]{8.2cm}
      \includegraphics[width=8.2cm]{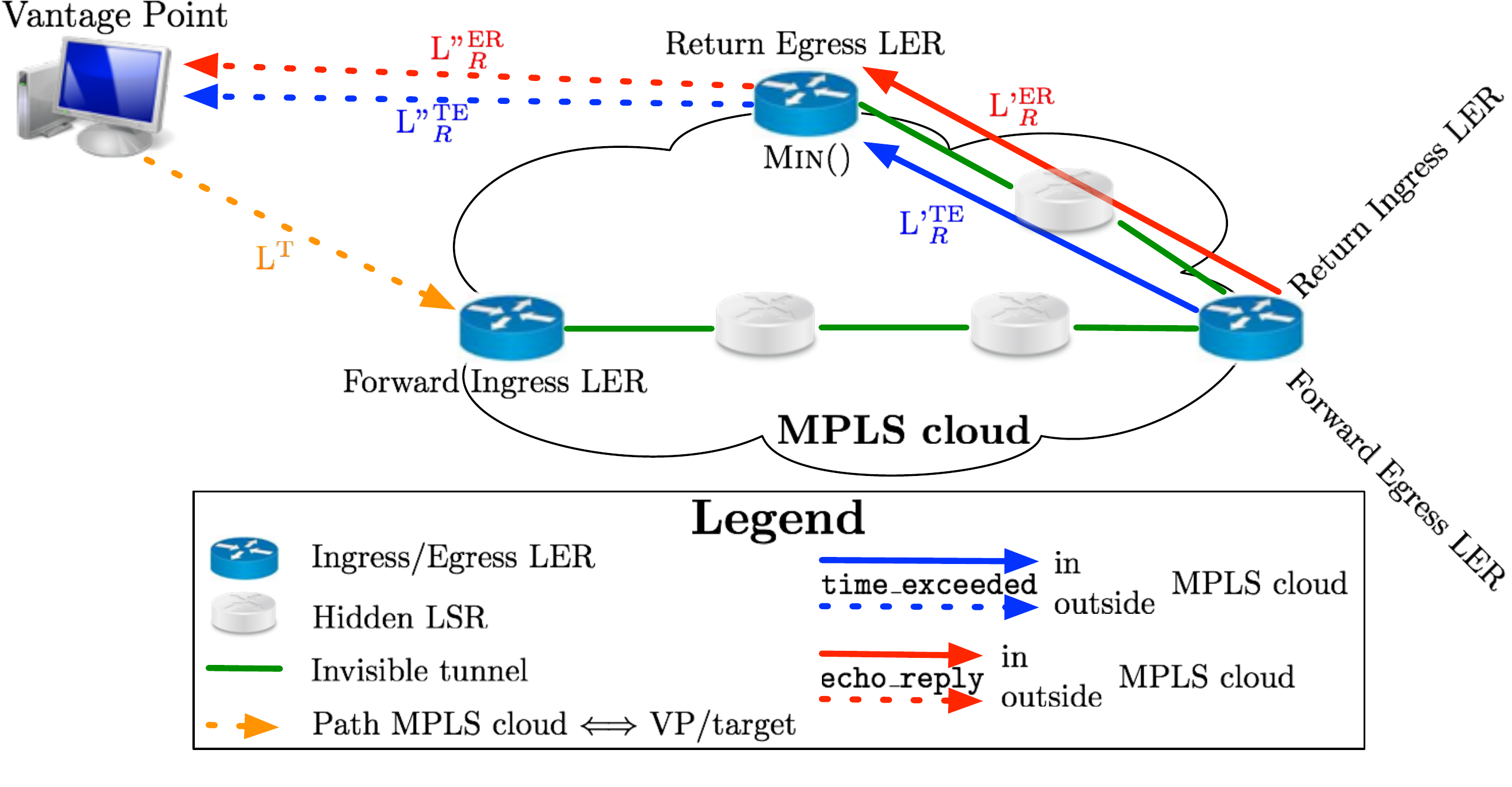}
      \caption{Invisible tunnels.}
      \label{tnt.trig.asym.invisible}
    \end{subfigure}
  \end{center}
  \caption{Indicators and triggers illustration for Implicit and Invisible
  tunnels. Notations L'$^{\textrm{x}}_y$ and L''$^{\textrm{x}}_y$ refer to a
  given sub-length of an ICMP packet \texttt{x} on the \texttt{y} path (y being
  the forward or return path and x being a \echoreply or \traceroute ICMP
  packet, see Fig.~\ref{background.dp.fig}). For example, L'$^{TE}_R$
  gives the return path of the \ttlexceeded within the MPLS cloud, while
  L''$^{TE}_R$ is the return path of the \ttlexceeded between the MPLS
  cloud and the vantage point. Consequently, we have \lter =
  L'$^{TE}_R$ +  L''$^{TE}_R$.}
  \label{tnt.trig.asym}
\end{figure*}

Due to the iBGP path heterogeneity (the IGP tie-break rule in particular), the
BGP return path taken by the ICMP \echoreply message can be different from the
BGP return path taken by the \ttlexceeded reply. This is illustrated in
Fig.~\ref{tnt.trig.asym.implicit} where the two return paths in blue and red can
differ even outside the AS (L''$^{TE}_R$ can be distinct from L''$^{ER}_R$). As
a result, and because it may differ at each intermediate hop, the \uturn
indicator does not necessarily follow exactly Eqn.~\ref{eqn.uturn}. A small
variation may then appear in practice. In particular, a value of $0$ can hide a
true Implicit hop.

For JunOS routers, the situation is quite different. It turns out that, by
default (i.e., without enabling the \texttt{icmp-tunneling} feature -- see
Appendix~\ref{appendix.p2mp} for details), these routers send \ttlexceeded
replies directly to the source, without forwarding them to the egress LER. The
\uturn indicator becomes then useless. Moreover, for routers having the JunOS
signature, the \uturn indicator and the \rtla trigger are computed in the same
way. Thus, to avoid any confusion, \tnt introduces an exception for such OS
signatures (line~\ref{exception} in Listing~\ref{tnt.trig.check_indicators}),
and first considers the difference as a trigger, and then falls back to an
indicator if the revelation fails (not shown in
Listing~\ref{tnt.overview.trace_tunnel} for clarity).
In addition, when \texttt{icmp-tunneling} is enabled, \ttlexceeded replies start
with a TTL of $254$, implying a bigger difference with \echorequest replies, as
it can be seen in Fig.~\ref{background.dp.fig}:
$\uturn(P_1)=$\ljerr-\ljter$=10-3=7$ instead of $6$ if $P_1$ runs a Cisco OS.

\begin{lstlisting}[escapeinside={(*}{*)}, label=tnt.trig.check_trigger,
caption=Pseudo-code for checking triggers]
code check_triggers(prev_hop, cur_hop, next_hop):
	#prev_hop and cur_hop must exist
	#duplicate IP checked on cur_hop and next_hop
	if (prev_hop == None or cur_hop == None or prev_hop == cur_hop)
		return None

	if (cur_hop == next_hop)(*\label{uhp}*)
		#invisible UHP tunnel
		return (*\dupip*)
	#retrieve path length from raw TTLs
	(*\lter*) = path_len(cur_hop.ttl_te)(*\label{dist1}*)
	(*\lerr*) = path_len(cur_hop.ttl_er)
	(*\lt*) = cur_hop.probe_ttl(*\label{dist2}*)

	if (sign_is_junOS(cur_hop))(*\label{finger}*)
		#for the JunOS signature
		if ((*\lter*) - (*\lerr*) (*$\geq$ \trtla*))(*\label{rtla}*)
			#invisible PHP tunnel with JunOS
			return (*\rtla*)
	else
	 	#for other signatures (raw TTLs are initialized the same)
		if ((*\lter*) - (*\lt*) (*$\geq$ \tfrpla*))(*\label{frpla}*)
			#invisible PHP tunnel with other known OS
			return (*\frpla*)

	return None
\end{lstlisting}

Indicators are MPLS passive pieces of evidence that can also prevent
\tnt from firing new probes (with the exception of LSE-TTL that is also a
trigger for Opaque tunnels).
On the contrary, triggers are active patterns suggesting the presence of
Invisible tunnels (both PHP and UHP) that could be revealed using additional
probing (see Sec.~\ref{tnt.revelation}).
Listing~\ref{tnt.trig.check_trigger} provides the pseudo-code for checking
triggers.

First, we look for potential Invisible UHP tunnel (line~\ref{uhp}). As explained
in Sec.~\ref{taxo}, Invisible UHP tunnels occur with Cisco routers
using IOS 15.2.  When receiving a packet with an IP-TTL of 1, the Egress LER
does not decrement the TTL but, rather, forwards it directly to the next hop.
Consequently, the Egress LER does not appear in the trace while, on the
contrary, the next hop (CE$_2$ in Fig.~\ref{background.dp.fig}) appears twice
(duplicate IP address in the trace output).

The two remaining triggers, \rtla (Return Tunnel Length
Analysis~\cite{mpls-invisible} -- see Table~\ref{tnt.trig.acronyms} for a
summary of acronyms used by \tnt) and \frpla (Forward/Return Path Length
Analysis~\cite{mpls-invisible}), work by using three path lengths, which are
\lter (the \ttlexceeded path length), \lerr (the \echoreply path length), and
\lt (the forward \traceroute path length). More precisely, \rtla is the
difference between the \ttlexceeded and the \echoreply return path lengths,
while \frpla is the difference between the forward and the return path lengths
(obtained based on \traceroute probe and reply messages). \tnt tries to capture
significative differences between these lengths to infer the presence of MPLS
tunnels, relying on two common practices of LSRs, in particular the EH,
developed in the previous subsection. Both triggers are based on the idea that
replies sent back to the vantage point are also likely to cross back the MPLS
cloud, which will apply the \minttl{IP-TTL}{LSE-TTL} operation at the EH of the
return tunnel. These triggers respectively infer the exact (\rtla) or
approximate (\frpla) return path length. Indeed, \frpla is subject to BGP path
asymmetry (and so, to false positives or negatives) in opposition to \rtla when
it applies (it may produce some false alarms but only due to ECMP). In the
absence of Invisible tunnel, we expect those triggers to have a value equal or
close to $0$. Indeed, in such a case, we should have L'$^{ER}_R =$ L'$^{TE}_F=$
L'$^{TE}_R=1$ if BGP does not interfere (see Fig.~\ref{tnt.trig.asym}).
Therefore, any significant deviation from this value is interpreted as the
potential presence of an Invisible MPLS cloud, and thus, brings \tnt to trigger
additional path revelation techniques (see Sec.~\ref{tnt.revelation}). In
practice (look at Fig.~\ref{tnt.trig.asym.invisible}), we expect to have
L'$^{ER}_R =$ L'$^{TE}_F = 1$ (due to the \textsc{Min} for the \echoreply return
tunnel and the pipe mode for the forward tunnel) while L'$^{TE}_R$ directly
provides the actual return tunnel length (with a value $\geq 1$). It is due to
the \textsc{Min} operation applied by the EH of the return tunnel, which selects
the LSE-TTL of the \ttlexceeded reply, and keeps the IP-TTL for the \echoreply
packet. Indeed, in the case of the \ttlexceeded message, the return Ingress LER
(i.e., the forward Egress LER) initializes the LSE-TTL with the same value as
the IP-TTL, meaning 255. For \echoreply packets, the IP-TTL is set to 64. \rtla
is not subject to any BGP asymmetry because we have L''$^{ER}_R =$ L''$^{TE}_R$,
i.e. BGP return paths have the same length. Indeed, the two messages use the
same physical path, the only difference being the \textsc{Min} operation applied
at the EH of the return tunnel, if any.

To check for those triggers, we first extract the three key distances thanks to
the reply IP-TTLs received by the vantage point (lines~\ref{dist1}
to~\ref{dist2} in Listing~\ref{tnt.trig.check_trigger}). As explained by
Vanaubel et al.~\cite{mpls-invisible}, \rtla only works with JunOS routers,
while \frpla is more generic. Therefore, prior to estimate the triggers, \tnt uses
network fingerprinting (see Sec.~\ref{background.fingerprinting}) to determine
the router brand of the potential Egress LER (line~\ref{finger} in
Listing~\ref{tnt.trig.check_trigger}).

In the presence of a JunOS hardware, \lter is compared to \lerr, as in case of
an Invisible tunnel, \lter is supposed to be greater than \lerr.
Indeed, with this routing platform, \ttlexceeded and \echoreply packets have
different initial TTL values (see Table~\ref{background.fingerprinting.table}),
and the \rtla trigger can exploit the TTL gap between those two kinds of
messages caused by the \minttl{IP-TTL}{LSE-TTL} behavior at the Egress LER (the
\lerr appears longer than \lter as the \textsc{Min} operation results in a
different pick). This difference represents the number of LSRs in the return
LSP, and is compared to a pre-defined threshold \trtla (line~\ref{rtla} in
Listing~\ref{tnt.trig.check_trigger}). This threshold (see Sec.~\ref{tnt_calib}
for the parameter calibration) filters all the LSPs shorter than the limit it
defines. In the case depicted in Fig.~\ref{background.dp.fig}:
\begin{eqnarray*}
  \rtla(PE_2) &=& L^{ER}_{R} - L^{TE}_R\\
              &=& L'^{ER}_R - L'^{TE}_R\\
              &=& 6 - 3\\
              &=& 3 .
\end{eqnarray*}

Indeed, for the \echoreply message, we have
{\small
\begin{eqnarray*}
\textrm{IP-TTL} & = & 64\\
		 & = & \textsc{Min}(\textrm{IP-TTL}=64,\textrm{LSE-TTL}=252) .
\end{eqnarray*}
}
instead of
{\small
\begin{eqnarray*}
\textrm{IP-TTL} & = & 252\\
		 & = & \textsc{Min}(\textrm{IP-TTL}=255,\textrm{LSE-TTL}=252) .
\end{eqnarray*}
}
for the \ttlexceeded reply. Note that an invisible shadow effect also
applies for \rtla after the Invisible tunnel, as the trigger will still be
positive for a few nodes after the egress LER.

\frpla is more generic and applies thus to any configuration. \frpla allows to
compare, at the AS granularity, the length distribution of forward (i.e., \lt)
and return paths (i.e., \lter). Return paths are expected to be longer than
forward ones, as the tunnel hops are not counted in the forward paths while they
are taken into account in the return paths (due to the \minttl{IP-TTL}{LSE-TTL}
behavior at the return Egress LER).Then, we can statistically analyze their
length difference and check if a shift appears (see Line~\ref{frpla} in
Listing~\ref{tnt.trig.check_trigger}). This is illustrated in
Fig.~\ref{background.dp.fig} (``Invisible PHP'') in which \lt is 3 while \lter
is equal to 6, leading so to an estimation of the return tunnel length of 3.  In
general, when no IP hops are hidden, we expect that the resulting distribution
will look like a normal distribution centered in 0 (i.e., forward and return
paths have, on average, a similar length). If we rather observe a significant
and generalized shift towards positive values, it means the AS makes probably
use of the \notpropagate option. In order to deal with path asymmetry, \tnt uses a
threshold, \tfrpla (see Sec.~\ref{tnt_calib} for calibrating this parameter),
greater than $0$ to avoid generating too much false positives (revelation
attempt with no tunnel). The \textsc{Min} effect also results in an invisible
shadow after the hidden LSP: $\frpla(CE_2)=2$ and $\frpla(CE_3)=1$, etc until
the situation returns to normal. Note that the \rtla and \frpla shadows are the
reasons why \tnt does not look for consecutive Invisible tunnels in a trace.
Finally, for Invisible UHP, one can observe that no \textsc{Min} shift applies
on the return path, as only the duplicate effect is visible.

Threshold calibration will be discussed in details in Sec.~\ref{tnt_calib}. The
optimal calibration can provide a 80/20 \% success/error rates (errors being due
to the BGP and ECMP noises). Moreover, the order in which \tnt considers
indicators and triggers, their codes, reflects their reliability, and so, their
respective success rates (and their resulting states): the lower the code (i.e.
the higher its priority), the more reliable (and higher the revelation success
rate). Thus, if a hop matches simultaneously multiple triggers (\rtla and \frpla
for example), it is tagged with the one having the highest priority (i.e., \rtla
in our example).

\begin{table}[!t]
{
\small
  \begin{center}
	\begin{tabular}{l|l|l}
	  \textbf{Acronym} & \textbf{Meaning} & \textbf{Usage}\\
	  \hline
	  \frpla & \stuff{F}orward/\stuff{R}eturn \stuff{P}ath \stuff{L}ength \stuff{A}nalysis & \multirow{2}{*}{Trigger}\\
	  \rtla  & \stuff{R}eturn \stuff{T}unnel \stuff{L}ength \stuff{A}nalysis & \\
	  \hline
	  \dpr  & \stuff{D}irect \stuff{P}ath \stuff{R}evelation & \multirow{2}{*}{Path Revelation}\\
	  \brpr & \stuff{B}ackward \stuff{R}ecursive \stuff{P}ath \stuff{Revelation} & \\
	\end{tabular}

  \end{center}
}
  \caption{Summary of acronyms used by \tnt.}
  \label{tnt.trig.acronyms}
\end{table}

\subsection{Hidden Tunnels Revelation}\label{tnt.revelation}
\begin{lstlisting}[escapeinside={(*}{*)}, label=tnt.revelation.reveal_tunnel,
caption=Pseudo-code for revealing Invisible tunnels]
state reveal_tunnel(ingress, egress, tun_code):
  #ingress and egress hops must exist
  if (ingress == None or egress == None)
    return None
  buddy_bit = False
  #standard (*\traceroute*) towards the candidate egress
  target = egress
  route = trace(REV_STARTING_TTL, target)(*\label{trace_egress}*)

  if (last_hop(route) != egress)
    #the target does not respond (revelation is not possible)
    return (*\tnotreached*)(*\label{target_not_reach}*)
  else if (ingress (*$\notin$*) route)
    #the forwarding path differs (revelation is not possible)
    return (*\ingnotfound*)(*\label{ing_not_found}*)
  else if (distance(ingress, egress, route) > 1)
    #path segment revelation with \dpr
    push_segment_to_revelation_stack(ingress, egress, route)
    return (*\dpr*)(*\label{dpr}*)
  else
    ttl = ingress.probe_ttl + 1
    revealed_ip = extract_hop(ttl, route)

    for iTR=0;;
      if (revealed_ip == target)
        if (tun_code != (*\dupip*) || buddy_bit)
          #no more progression in the revelation
          break
        else
          #try with the buddy for the (*\dupip*) trigger
          target = buddy(revealed_ip)(*\label{buddy}*)
          buddy_bit = True
      else
        #a new hop has been revealed
        iTR++
        push_hop_to_revelation_stack(revealed_ip)
        target = revealed_ip
        buddy_bit = False

      revealed_ip = traceHop(ttl, target)

  if (iTR == 0)
    #no revelation (fail)
    return NOTHING_TO_REVEAL(*\label{nothing_to_reveal}*)
  if (iTR == 1)
    #single hop revealed LSP (\dpr (*$\approx$*) \brpr)
    return 1HOP_LSP(*\label{unknown}*)
  else
    #hop by hop revelation with \brpr
    return (*\brpr*)(*\label{brpr}*)
\end{lstlisting}

Listing~\ref{tnt.revelation.reveal_tunnel} offers a simplified view of the \tnt
tunnel revelation. The first step consists in launching a standard \traceroute
towards the candidate Egress\footnote{We use the term \dfn{candidate} as, at
this point, we are not completely sure an MPLS tunnel is hidden there.}
(line~\ref{trace_egress} in Listing~\ref{tnt.revelation.reveal_tunnel}).
\texttt{REV\_STARTING\_TTL} is the starting TTL used for the revelation, which
corresponds to 2 hops before the candidate Ingress hop, by default.  During this
first attempt, \tnt may fail to reach the candidate Egress
(line~\ref{target_not_reach}), and/or the candidate Ingress
(line~\ref{ing_not_found}) when collecting the active data. Otherwise, \tnt may
reveal a tunnel and four additional output states can arise:
\begin{itemize}
  \item an LSP composed of at least 2 LSRs is revealed in the first trace
  towards the Egress (line~\ref{dpr} -- \dpr, Direct Path
  Revelation~\cite{mpls-invisible} -- Table~\ref{tnt.trig.acronyms}
  for a summary of acronyms used by \tnt);
  \item an LSP having more than one LSR is revealed using several iterations
  (line~\ref{brpr} -- \brpr, Backward Recursive Path
  Revelation~\cite{mpls-invisible}).
  \item nothing is revealed, the candidate Ingress and Egress are still
  consecutive IP addresses in the trace towards the candidate Egress
  (line~\ref{nothing_to_reveal});
  \item a single-hop LSP is revealed (line~\ref{unknown}) although
  several iterations have been tried:  \dpr and \brpr cannot be distinguished for
  one hop LSPs.
\end{itemize}

With the default UHP configuration on Cisco IOS 15.2, an additional test, called
\dfn{buddy} (line~\ref{buddy}), is required to retrieve the outgoing IP
interface of the Egress LER (the right interface, in green, on PE$_2$ in
Fig.~\ref{background.dp.fig}), and thanks to its retrieval, force replies from
its incoming IP interface (the left one, in red, on PE$_2$ in
Fig.~\ref{background.dp.fig}). The \texttt{buddy()} function assumes a
point-to-point connection between the Egress LER and the next hop (IP addresses
on this point-to-point link are called \dfn{buddies}). In most cases, the
corresponding IP addresses belong to a /31 or a /30
prefix~\cite{merlin-degree,treenet}.  Note that according to the IP address
submitted to \texttt{buddy()}, this function may require additional probing to
infer the correct prefix in use.
Besides, specific UDP probing is necessary in order to provoke
\dstunreach messages. Such error messages, as \ttlexceeded ones, enable to get
the incoming interface of the targeted router (instead of \echoreply ones that
are indexed with the target IP).

\dpr (Direct Path Revelation) works when there is no MPLS tunneling for internal
IGP prefixes other than loopback addresses, i.e., the traffic to internal IP
prefixes is not MPLS encapsulated (default Juniper configuration but can also be
easily configured on Cisco devices -- see Sec.~\ref{background.cp}) .
With PHP, \brpr (Backward Recursive Path Revelation) works because the target
(PE$_2$.left on Fig.~\ref{background.dp.fig}) belongs to a prefix being also
advertised by the PH. Thus, the probe is popped one hop before the PH (P$_3$ on
Fig.~\ref{background.dp.fig}), and it appears in the trace towards the Egress
incoming IP interface, e.g., PE$_2$.left on Fig.~\ref{background.dp.fig}. \brpr
is then applied recursively on the newly discovered interface until no new IP
address is revealed. \brpr works also natively with UHP on IOS 12.4 (i.e.,
without the \texttt{buddy()} function), for the same reason as for PHP: the
prefix is local and shifts the end of the tunnel one hop before and, in this
implementation, the EH replies directly. On the contrary, \tnt needs to use the
\texttt{buddy()} function at each step for IOS 15.2 enabling UHP, because the EH
silently forwards the packet one hop ahead. Vanaubel et al.~\cite{mpls-invisible}
provides more details on \dpr and \brpr.

\subsection{Reproducibility and Practical BGP Configurations}\label{tnt.validation}
We use the  GNS3  emulation environment for several purposes.
First, we aim at verifying that the inference assumptions we considered in the
wild are correct and reproducible in a controlled environment. Second, some of
the phenomena we exploit to reveal tunnels in the wild have been directly
discovered in our testbed. Indeed, using our testbed we reverse-engineered the
TTL processing (considering many MPLS configurations, we study the \pop
operation in particular) of some common OSes used by many real routers. Finally,
it is also useful for debugging \tnt to test its features in this controllable
environment. Generally speaking, we aim at reproducing with GNS3 all common
behaviors observed in the wild, and, on the opposite, we also expect
to encounter in the wild all basic behaviors (based on standard MPLS and BGP
configurations) we build and setup within GNS3.

In practice, we have considered four distinct router OSes: two  Cisco
standard IOS (12.4 and 15.2), and  two virtualized versions of
JunOS (Olive and VMX, the only Juniper OSes we succeeded to emulate within
GNS3). We envision in a near future to also test the IOS XR and some other
Juniper OSes, if possible, but we believe that our tests are already
representative enough of most behaviors existing in the wild.

In our emulations, topologies (see Fig.~\ref{background.dp.fig}) are configured
as follows.  We assumed that LERs are AS Provider-Edge (PE) routers, i.e., AS
border routers of the ISP running (e)BGP sessions.   Two main configurations are
then possible to enable transit tunneling at the edges. Either the BGP next-hop
can be the loopback IP address of the PE itself (with \texttt{next hop self}
command), or it belongs to the eBGP neighbor -- and in that case the connected
subnet or the IP address should be redistributed in the ISP. In both cases,
there exists a LDP mapping, at each Ingress LER and for any transit forwarding
equivalent class (FEC) between the BGP next-hop, the IGP next-hop, and the local
MPLS label to be pushed. According to the configuration at the Egress LER, when
the Ingress LER is in pipe mode (see Sec.~\ref{background.dp.push}), distinct
kinds of tunnels emerge: Opaque, UHP Invisible, or PHP Invisible.

We consider the simplest possible configurations, i.e., homogeneous in terms of OS
and MPLS+BGP configurations. They are consistent and symmetric MPLS
configurations both in terms of signaling (LDP with the independent  model using
all IGP connected prefix -- Cisco default mode -- xor the ordered model using
only loopback addresses -- Juniper default mode)\footnote{See
Sec.~\ref{background.cp}} and the propagation operation in use (pipe xor
uniform)\footnote{See Sec.~\ref{background.dp.push}} at the domain scale.
Using heterogeneous configurations, we discovered many intriguing corner cases
that are discussed in Appendix~\ref{appendix.corner}. Some of them may result in
incorrect TTL processing and other in hiding even more the tunnel to \tnt. In some rare
cases, only the Brute Force option of \tnt is able to fire the path revelation
that exposes tunnels.

The BGP configuration is also standard: the Egress LER enables the next-hop-self
feature and so the transit traffic is tunneled via this IP address. All LSRs
also have a global IGP routing table thanks to a route reflector (they can
answer natively to ping requests) or a redistribution in the IGP routing control
plane. The AS scale BGP prefix is advertised using a global aggregation and the
BGP inter-domain link is addressed by the neighbor but can be redistributed in
the IGP as a connected one.

Opaque tunnels show up when enabling the \texttt{neighbor <IP> ebgp-multihop
<\#hops>} command towards the BGP neighbor whose IP address is redistributed
statically in the IGP. \dpr works also with Cisco IOS when enabling the
\texttt{mpls ldp label allocate global host-routes} command. Eventually, the
command \texttt{mpls ldp explicit-null [for prefix-acl]} allows for revealing
UHP tunnels without the use of the buddy.

Table~\ref{tnt.alltech.table} provides a summary of \tnt capacities considering
several MPLS usages.  In particular, it provides many information about the way
\tnt is able to collect information about tunnels in default cases (i.e.,
standard configurations). For example, it shows that \tnt is able to
discriminate between Cisco Invisible UHP and PHP tunnels while it is not the
case for Juniper routers (the \CheckedBox~is not colored). Indeed, for both
UHP/PHP Juniper configurations, the trigger and the revelation methods are the
same (RTLA and \dpr). Moreover, we also show when our basic set of techniques
need to be extended for enabling revelation and distinction among different
classes: we use the symbol $++$ to enforce these new requirements. In
particular, for revealing UHP Cisco tunnels, \tnt extends \brpr with additional
\texttt{buddy()} function and UDP probing.

\begin{table}[!t]
\begin {tabular} {l|c|c|c}
     Configurations & Pop & Cisco iOS15.2 & Juniper VMX \\
   \hline
   P2P circuits     & PHP & FRPLA, \brpr & RTLA, \dpr \\
   (e.g. LDP or     & UHP & DUP\_IP, \brpr++ & RTLA, \dpr \\
   RSVP-TE tunnels) &    & \color{green}{\CheckedBox} & \CheckedBox \\
   \hline
   P2MP overlays & PHP & LSE-TTL, - & RTLA++, - \\
   (e.g. VPRN: CsC or & UHP & LSE-TTL++, - & N/A \\
   VPN BGP-MPLS) &  &  \color{green}{\XBox} & \XBox\\
\end {tabular}
\caption{\tnt revelation (\CheckedBox) and detection (\XBox) capacities
according to the OS and the MPLS tunnel flavor (i.e. the MPLS L3 tunneling
underlying technologies for a given usage). In particular, this table provides
the default indicator/trigger and the default path revelation method (when it
applies).}
\label{tnt.alltech.table}
\end{table}

Another kind of MPLS technology that conduct to surprising patterns is
VPRN tunnels. They also lead to Opaque tunnels for Cisco routers
and twisted traces with Juniper routers. 

As shown by Table~\ref{tnt.alltech.table}, \tnt is able to reveal the content of
all basic point-to-point (P2P) MPLS circuits (i.e., LDP or RSP-TE tunnels).
However, things are not as easy when point-to-multipoint (P2MP) tunnels join the
party (second part of Table~\ref{tnt.alltech.table}), leading typically to
Opaque tunnels.  In such a situation, \tnt is only able to detect their presence
but without revealing their content. Indeed, such tunnels may result from
various network configurations, e.g., heterogeneous routing devices (a
combination of Juniper and Cisco devices in particular), specific BGP edge
configurations as already introduced, or VPRN (Virtual Private Routed
Network)~\cite{rfc2764}. As for the two former configurations, the Opaque
pattern arises with VPRN because of an abrupt tunnel ending, i.e., the LSP ends
without a standard ending label (Implicit or Explicit NULL). Indeed, the last
hop towards the Egress contains at least one label (two with UHP, the top label
then being 0). The inner label is used to identify the VPN and the associated
VRF containing routes. By definition, the VPN label value is neither Explicit
NULL nor Implicit NULL. Upon receiving this non-terminating label, the Egress
behaves as if the tunnel did not end in a controlled fashion.

Using our GNS-3 platform, it appears that VPRN content cannot be revealed with
\tnt, while other Opaque tunnels configurations (i.e., routing devices
heterogeneity, BGP edge configuration) can.  The mechanism behind the absence of
content revelation can be explained by the IP address collected by \tnt from the
source IP field in the ICMP reply.  Usually, the collected address is the one
assigned to the physical incoming interface of the Egress PE. In the VPRN case, the
collected IP is the one assigned to the interface on which the VRF is attached.
In practice, this corresponds to the outgoing interface towards the VPN at
the customer's side. Said otherwise, \tnt collects the outgoing address instead
of the incoming one. Because the incoming address is the only one that
enables a successful revelation, this type of Opaque tunnels cannot be revealed
yet.  Table~\ref{tnt_quantif.results.tab} in Sec.~\ref{tnt_quantif} will show
that, in the wild, VPRNs are the most prevalent case of Opaque tunnels.

Whatever the kind of probes sent to or through the VPRN, the IP address visible
to \tnt (or \texttt{traceroute} in general) is the outgoing address. Despite
its expired TTL, it is likely that the probe arriving on the Egress PE will be
pushed to the VRF of the VPN and its associated interface before generating the error
message (the VPN being identified with the MPLS label contained in the packet).
Then, the interface where the packet actually expires is the one associated to
the VRF. However, as shown in Table~\ref{tnt.alltech.table}, we are able to
distinguish UHP and PHP configurations (thanks to LSE-TTL++), because the bottom
label is equal to $255$ for UHP and lower with PHP.

With Juniper VPN, there is no Opaque indicator resulting from VPRN or any other
configurations. A first explanation is that Juniper routers, on the contrary to
the independent mode enabled by default with Cisco routers, do not inject the
whole IGP in LDP, but only their loopback address using the ordered mode (see
Sec.~\ref{background}). This mode limits the probability to face a
non-controlled tunnel ending. However, with VPRN configurations, a Juniper
Egress LER deals with the same packet level situation as with Cisco routers. Up
to the end of the tunnel, the packet is still MPLS encapsulated with the
end-to-end VPN non terminating label at the bottom of the stack. Juniper routers
do not, however, produce an Opaque indicator in that situation. Indeed, packets
destined to the VPN are handled in a specific way with Juniper devices: they are
IP packets forwarded directly to the next-hop without looking at or manipulating
the IP-TTL whatever its value.

The outcome of such a sliding packet is twofold. Firstly, the Egress hop is
hidden in the transit trace, as with Cisco UHP but without the duplicated IP.
Secondly, when performing a direct trace (even with UDP) targeting the first
address of the path within the VPN, i.e the IP interface of the Egress LER
belonging to the VPN, one can see that this address and its buddy appear in the
wrong order. Indeed, in the trace, the two addresses are switched, meaning that
the CE IP address appears before the Egress one. Being forwarded without
inspecting the IP-TTL, probes targeting IP addresses belonging to the VPN are
automatically forwarded to the CE router, where they expire. The next probe,
having a greater TTL, follows the same path as the one before, but can be
forwarded back to the Egress LER by the CE router before expiring. This loop
results in the two addresses being switched regarding their actual location in
the path. Finally, one can infer the loop because two additional artifacts
compared to RTLA (RTLA++) are visible: the TTL that deviates from its monotony and
subsequent IP addresses also raise alarms due to potential conflicting
allocation.

Additional details on the validation through GNS3 emulation can be found in
the Appendix, at the end of this paper. All topologies and scripts developed are
available for reproducibility.\textsuperscript{\ref{footnote.web}}


\section{\tnt Calibration and Probing Cost}\label{tnt_calib}
Sec.~\ref{tnt} shows that \tnt relies mainly on four parameters when looking for
tunnels indicators or triggers: \tlsettl for Opaque tunnels, \tuturn for
Implicit tunnels, and \trtla and \tfrpla for PHP/UHP Invisible tunnels.  This
section aims at calibrating those parameters (Sec.~\ref{tnt_calib.calibration}), as well
as evaluating the probing cost associated to \tnt (Sec.~\ref{tnt_calib.cost}).

\subsection{Measurement Setup}\label{tnt_calib.setup}
We deployed \tnt on three vantage points (VPs) over the Archipelago
infrastructure~\cite{ark}.  VPs were located in Europe (Belgium), North America
(San Diego), and Asia (Tokyo).

\tnt was run on April \nth{6}, 2018 towards a set of 10,000 destinations
(randomly chosen among the whole set of Archipelago destinations list).  Each VP
had its own list of destinations, without any overlapping.

\begin{figure}[!t]
  \begin{center}
    \includegraphics[width=6cm]{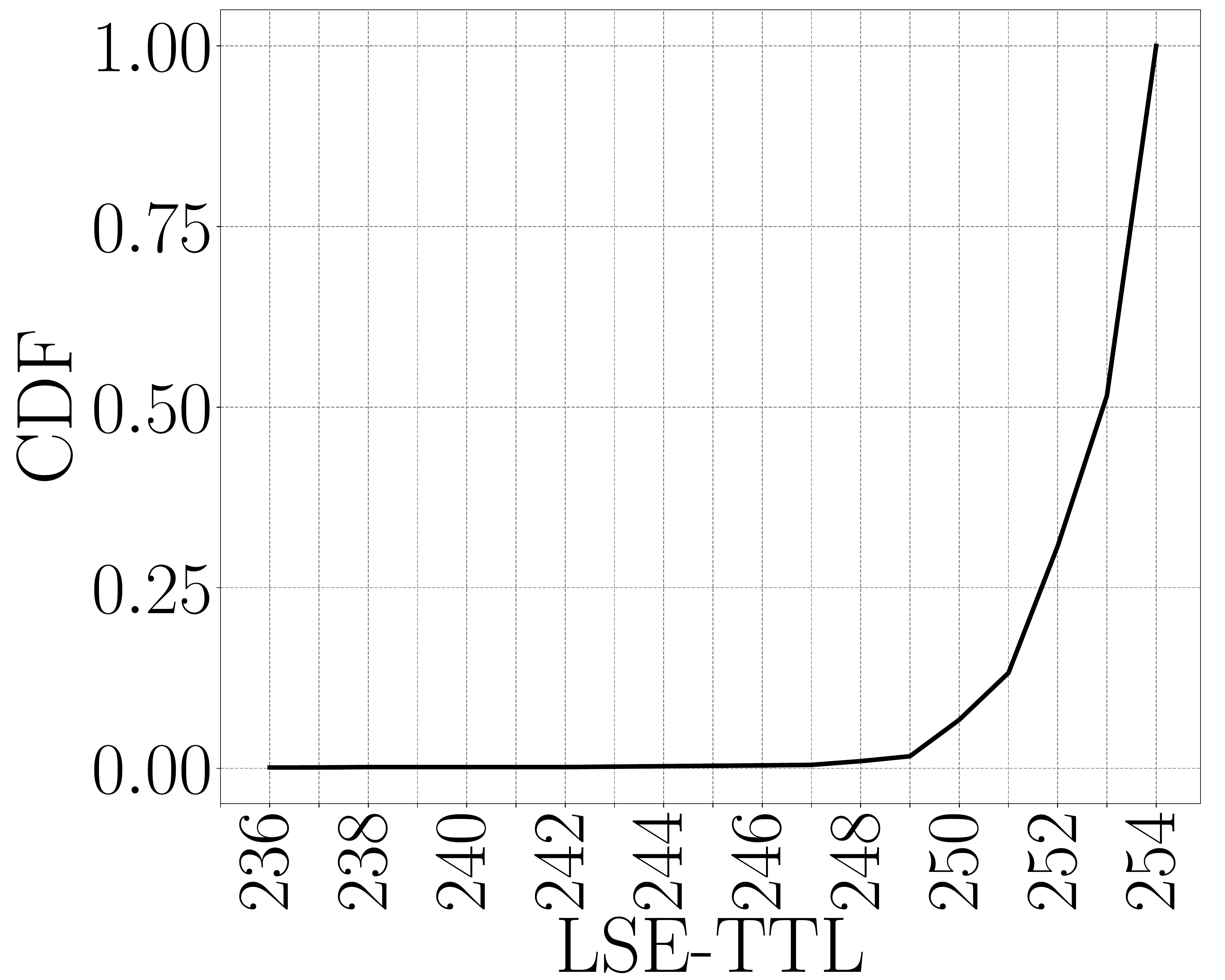}
  \end{center}
  \caption{Distribution of abnormal LSE-TTL values received at vantage points}
  \label{tnt_calib.setup.lsettl}
\end{figure}

From indicators and triggers described in Sec.~\ref{tnt.trig} (see
Listing~\ref{tnt.trig.check_indicators} and~\ref{tnt.trig.check_trigger}), it is
obvious that \uturn is equivalent to \rtla for Juniper routers. However, the
\tuturn will not have the same value than \trtla. \tuturn$=0$ by design as any
difference between \echoreply and \ttlexceeded replies for the Cisco router
signature indicates LSE-/IP-TTL shifting. In practice, we reinforce the
condition by looking for at least two consecutive hops having a cumulated \uturn
$\geq 3$. Besides, we have observed that abnormal\footnote{Abnormal here means
``different from 1'' which is the LSE-TTL value that should be obtained in ICMP
\ttlexceeded messages.} LSE-TTL values oscillate between 236 and 254, the
main proportion being located between 250 and 254, as shown in Fig.~\ref{tnt_calib.setup.lsettl}.  It suggests thus that, in
the majority of the cases, Opaque tunnels are rather short. Consequently, a
value of 236 for \tlsettl would be enough for detecting the presence of an
Opaque tunnel and launching additional measurements for possibly revealing its content.

For our tests, we varied \trtla and \tfrpla between 0 and 4. A full measurement
campaign was launched for each pair of parameter value (thus, a total of 25
measurement runs). Moreover for each pair, if no trigger is pulled, a so called
brute force revelation is undertaken: \dpr/BRPR  are launched (with the use of
the buddy if required). This brute force data is used as a basis to evaluate
the quality and cost of each threshold value.

\subsection{Calibration}\label{tnt_calib.calibration}
\begin{figure}[!t]
  \begin{center}
    \includegraphics[width=6cm]{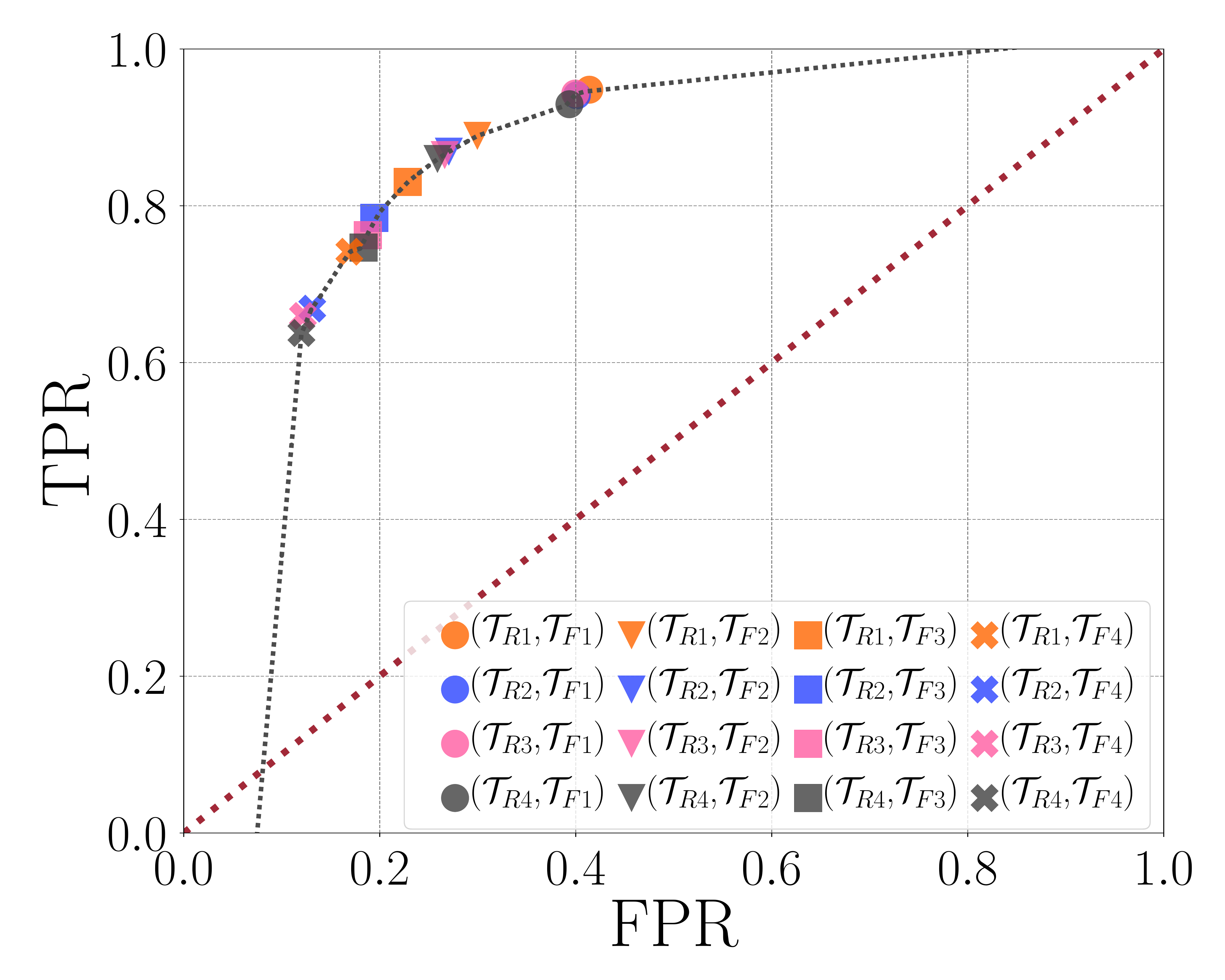}
  \end{center}
  \caption{Receiver operating characteristic (ROC) curve providing the
  efficiency of \tnt according to values for Invisible tunnels parameters.
  $\mathcal{T}_{R_x}$ refers to \trtla with the value $x$, while
  $\mathcal{T}_{F_y}$ to \tfrpla with the value $y$.}
  \label{tnt_calib.calibration.roc}
\end{figure}

With the help of well calibrated thresholds, the results associated to \frpla
and \rtla triggers allows for a binary classification. These triggers provide a
prediction, while the results of additional probing gives the true facts when
some conditions apply (see resulting states of
Listing~\ref{tnt.revelation.reveal_tunnel}), i.e., being or not a tunnel. With
that in mind, one can assess the performance of \frpla and \rtla triggers
through the analysis of True Positive Rate (TPR) and False Positive Rate (FPR):
we plot the results on a Receiver Operating Characteristic (ROC) curve in
Fig.~\ref{tnt_calib.calibration.roc}. We define TPR as the ratio of \tnt success
to the number of links being actually MPLS tunnels (having a length greater than
1): \tnt triggers additional probing and actually reveals Invisible tunnels (we
have $TPR+FNR=1$, i.e., when adding to False Negative Rate, we obtain all links
being long enough tunnels). FPR is defined as the ratio of \tnt failure to the
amount of standard IP links: it triggers for additional probing but without
revealing anything (we have $FPR+TNR=1$, i.e., when adding to True Negative
Rate, we obtain all IP links without tunnels). Here, our brute force data gives
the ground data that we consider reliable (i.e., revelation is fired at each hop
and if nothing is revealed, we consider that there is no tunnel -- we do not
consider inconclusive cases where we obtain states \ingnotfound or \tnotreached
-- see Listing~\ref{tnt.revelation.reveal_tunnel}).  The ROC curve is obtained
by varying the \trtla and \tfrpla parameters between 0 and 4. The red dotted
diagonal provides the separation between positive results for \tnt (above part
of the graph) and negative results (below part of the graph). Finally, the black
dotted line is the interpolation of measurement results (at the exception of
$\mathcal{T}_{R_0}$ values which appear as being outliers, as expected).

We observe that the results are essentially positive for \tnt.  Some results,
between ($\mathcal{T}_{R_1}$, $\mathcal{T}_{F_3}$) and ($\mathcal{T}_{R_2}$,
$\mathcal{T}_{F_3}$), are even reasonably close to the perfect classification
(upper left corner) and, thus, are considered as the best choice for defining
our thresholds \trtla and \tfrpla. We obtain a compromise close to 80\%-20\%:
while we expect to reveal at least 80\% of existing tunnels (MPLS links), \tnt
has a controlled overhead of 20\%, i.e., it fires useless additional probing for
an average limited to two actual IP links on ten.

\subsection{Probing Cost}\label{tnt_calib.cost}
\begin{figure}[!t]
  \begin{center}
    \includegraphics[width=6.5cm]{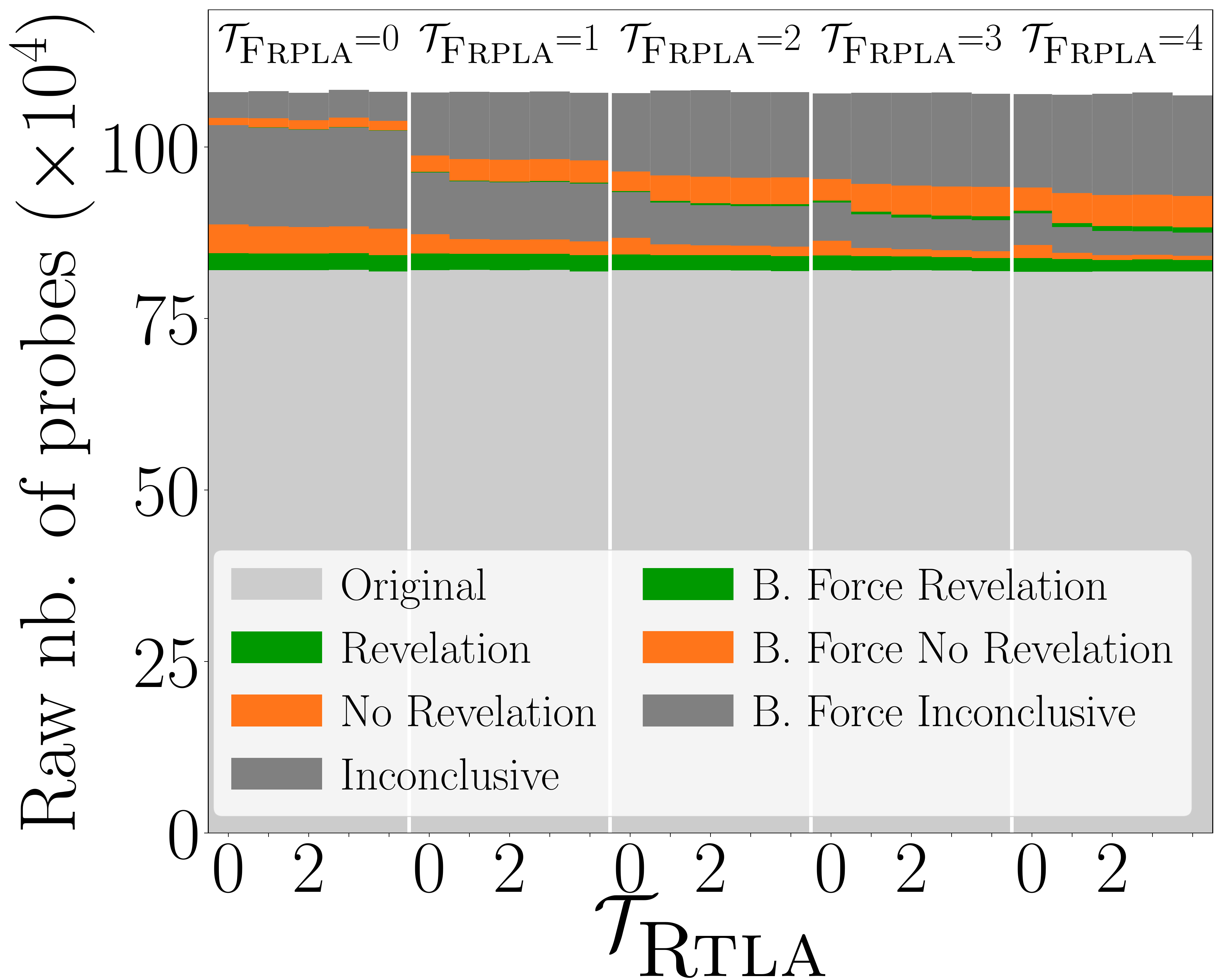}
  \end{center}
  \caption{Probing cost associated to \tnt according to \tfrpla and \trtla thresholds.}
  \label{tnt_calib.cost.fig}
\end{figure}

Fig.~\ref{tnt_calib.cost.fig} illustrates the probing cost associated to \tnt.
In particular, it focuses on additional measurements triggered by \rtla or
\frpla for revealing Invisible tunnels. The light grey zone (labeled as
``Original'' on Fig.~\ref{tnt_calib.cost.fig}) corresponds to probes associated
to standard \traceroute. The green, orange, and dark grey zones correspond to
probes sent when additional measurements are triggered by \rtla or \frpla. In
particular, the green zone corresponds to additional measurements that were able
to reveal the content of an Invisible tunnel. On the contrary, the orange zone
refers to additional measurements that failed, i.e., no Invisible tunnel content
was revealed.  Finally, the dark grey zone refers to inconclusive revelation:
the trigger has led to additional measurements but \tnt was unable to reach the
potential Egress LER (i.e., the IP address that engaged the trigger --
\texttt{cur\_hop} in Listing~\ref{tnt.overview.trace_tunnel} -- generally due to
unresponsive IP interface) or \tnt was unable to reach again the candidate
Ingress LER (i.e., \texttt{prev\_hop} in
Listing~\ref{tnt.overview.trace_tunnel}) because the destination has changed
(ECMP or BGP routing noises).

If the amount of probes sent for actually revealing the content of an Invisible
tunnel remains almost stable whatever the values for \tfrpla and \trtla are,
one can observe a very slow decrease meaning that there are less revealed
tunnels for high values.  Further, the additional traffic generated by erroneous
trigger (orange) or by inconclusive revelation (dark grey) clearly decreases while
\tfrpla increases.  This result is aligned with Sec.~\ref{tnt_calib.calibration}
in which the best values for \tfrpla are between 2 and 3. Note that \frpla is
more generic but less reliable than other triggers.
On the contrary, the \trtla threshold has a minor effect on the amount of probes
sent because it is more specific and more reliable.

Hatched zones (orange, dark grey, and green) correspond to the amount of probes
sent using brute force.
First, on the contrary to normal behavior (i.e., revelation launched according
to triggers), the amount of probes sent increases with \tfrpla (the impact of
\trtla is quite negligible), as well as the amount of inconclusive revelation.
Second, the amount of probes having revealed an Invisible tunnel is low compared
to standard behavior.

Generally speaking, one can observe that the overhead of \tnt is quite limited
compared to a basic active campaign and considering the information gathered. In
particular, if using correct parameters to limit both useless probes and missed
tunnels (e.g., $\mathcal{T}_{R_1}$, $\mathcal{T}_{F_3}$), our tool generates
less than 10\% of additional probing compared to the underlying campaign for
reaching a satisfying compromise where 80\% of tunnels are revealed.


\section{Tunnels Quantification With \tnt}\label{tnt_quantif}
This section aims at discussing how \tnt and its features behave in the wild
Internet. In particular, it analyzes the success rate of each indicator and
trigger with respect to possible revelation techniques.
Sec.~\ref{tnt_quantif.setup} describes the measurement setup, while
Sec.~\ref{tnt_quantif.results} discusses the results obtained.

\subsection{Measurement Setup}\label{tnt_quantif.setup}
We deployed \tnt on the Archipelago infrastructure~\cite{ark} on April \nth{23},
2018 with parameters \tfrpla fixed to $3$ and \trtla to $1$, according to
results discussed in Sec.~\ref{tnt_calib.calibration}.

\tnt has been deployed over 28 vantage points, scattered all around the world:
Europe (9), North America (11), South America (1), Asia (4), and
Australia (3).  The overall set of destinations, nearly 2,800,000 IP addresses, is
inherited from the Archipelago dataset and spread over the 28 vantage
points to speed up the probing process.

A total of 522,049 distinct IP addresses (excluding \traceroute targets) has
been collected, with 28,350 being non publicly routable addresses (and thus
excluded from our dataset). Each collected routable IP address has been
pinged, only once per vantage point, allowing us to collect additional data for
fingerprinting (see Sec.~\ref{background.fingerprinting}).  Our dataset and our
post-processing scripts are freely available.\textsuperscript{\ref{footnote.web}}

\subsection{Results}\label{tnt_quantif.results}
\begin{table}[!t]
  \begin{center}
     \begin{tabular}{ll|c|c|c}
       \multicolumn{2}{c|}{\multirow{2}{*}{\textbf{Status}}} & \multicolumn{3}{c}{\textbf{\# probes}}\\
        & & \traceroute & \ping & \texttt{buddy}\\
      \hline
      \multicolumn{2}{c|}{original} & 63,559,385 & 7,109,075 & $-$\\
      \hline
      \multirow{4}{*}{\rotatebox{90}{attempt}} & revealed      & 2,190,275
      & 206,842 & 19,181\\
                               & no revelation & 1,640,224      & $-$     & 556\\
                               & \tnotreached & 4,174,404 & $-$     & 9,888\\
                               & \ingnotfound  & 1,790,900 & $-$     & 7,326\\
    \end{tabular}
  \end{center}
  \caption{Raw number of probes sent by \tnt over the set of 28 vantage
  points.}
  \label{tnt_quantif.results.probes}
\end{table}

Table~\ref{tnt_quantif.results.probes} provides the amount of probes sent by
\traceroute-like probing in \tnt, \texttt{ping}, and buddy bit exploration. The
row ``original'' refers to standard \traceroute based revelation (i.e.,
nothing to reveal, Explicit, or Implicit tunnels).

The main results from Table~\ref{tnt_quantif.results.probes} is the amount of
probes involved in inconclusive revelation, split between \tnotreached (\tnt was
unable to reach the potential Egress LER) and \ingnotfound (\tnt did not cross
the potential Ingress LER). In particular, \tnotreached involved twice more
probes than revealed tunnels. Those particular inconclusive revelations might be
explained by ICMP rate limiting between the \traceroute probe and additional
probing (both \texttt{ping} and BRPR/DPR). Another explanation is that those
potential Egress LERs respond to initial \traceroute with an IP address that is
not globally announced. As such, additional probing following \traceroute will
fail as no route is available to reach them.

{\setlength{\doublerulesep}{0pt}
\setlength{\tabcolsep}{5pt}
\begin{table*}[!t]
  \begin{center}
    \begin{tabular}{Sl|Sl|ScScScSc|Sc}
    \multirow{2}{*}{\textbf{Tunnel Type}} & \multirow{2}{*}{\textbf{{\color{orange}Indicator}}/{\textbf{\color{red}Trigger}}} & \multicolumn{4}{c|}{\textbf{Revelation Technique}} & \multirow{2}{*}{\textbf{\# Tunnels}}\\
          & & DPR & BRPR & \unknown & Mix & \\
    \hline
    Explicit & {\color{orange}LSE headers} & - & - & - & - & 150,036\\
    \hline
    \multirow{2}{*}{Implicit} & {\color{orange}qTTL} & - & - & - & - & 2,689\\
    						  & {\color{orange}\uturn} & - & - & - & - & 7,216\\
    \hline
    Opaque & {\color{orange}LSE-TTL} & 22 & 17 & 43 & - & 3,346\\
    \hline
    \multirow{2}{*}{Invisible PHP} & {\color{red}\rtla}  & 11,268 & 1,191 & 2,595 & 279   & 15,333\\
    							   & {\color{red}\frpla} & 5,903	 & 2,555 & 3,260 & 1,012 & 12,730\\
    \hline
    Invisible UHP & {\color{red}\dupip} & 1,609 & 1,531 & 686 & 296 & 4,122\\
    \hline\hline\hline\hline
    \multicolumn{2}{c|}{\textbf{Total}} & 18,802 & 5,294 & 6,584 & 1,587 & 195,525\\
    \end{tabular}
  \end{center}
  \caption{Raw number of tunnels discovered by \tnt per tunnel type (see
  Sec.~\ref{taxo}). Color code for indicators/triggers is identical
  to Fig.~\ref{background.dp.fig}. No additional revelation technique is
  necessary for Explicit and Implicit tunnels.}
  \label{tnt_quantif.results.tab}
\end{table*}
}

Table~\ref{tnt_quantif.results.tab} provides the number of MPLS tunnels
discovered by \tnt, per tunnel type as indicated in the first column. The
indicators/triggers are provided, as well as the additional revelation technique
used. Without any surprise, Explicit tunnels are the most present category
(76\% of tunnels discovered).

Implicit tunnels represent 5\% of the whole dataset, with the \uturn indicator
providing more results than qTTL. However, those results must be taken with care
as \uturn is subject to false positive (implicit \uturn tunnels are likely to be
overestimated because of possible confusion with RTLA for Juniper routers),
while qTTL is much more reliable~\cite{mpls-tma}. Compared to previous works, it
is clear that is class is not as prevalent as expected at the time (because both
we correct and improve our methodology and the RFC4950 is likely to be more and
more deployed).

Opaque tunnels are less prevalent (1.7\% of tunnels discovered).  This is
somewhat expected as Opaque tunnels are the results of particular label
distribution within Cisco MPLS clouds. This confirms previous empirical
results~\cite[Sec.~7.2]{mpls-ccr}.  It is also worth noticing that additional
revelation techniques (DPR or BRPR) does not perform well with such tunnels
(content of 98\% of Opaque tunnels cannot be revealed). Indeed, as already
discussed earlier, this result can be explained because Carrier-of-Carriers
VPN~\cite{rfc4364} or similar VPRN are not possible to reveal but can only be
detected. We deduce from our campaign results that the vast majority of Opaque
tunnels seems to arise from Cisco VPRN.\footnote{In this paper, \tnt does not
look for Juniper VPRN as its indicator, RTLA++, is less reliable (See
Sec.~\ref{tnt.validation}).}

The proportion of Invisible tunnels is not negligible (16\% of tunnels in our
dataset). Those measurements clearly contradicts our previous work suggesting
that Invisible tunnels were probably 40 to 50 times less numerous than Explicit
ones~\cite[Sec.~8]{mpls-ccr}. More precisely, Invisible PHP is the most
prominent configuration (87\% of Invisible tunnels belongs to the Invisible PHP
category), confirming so our past survey~\cite{mpls-invisible}. \rtla appears as
being the most efficient trigger. This is partially due to the order of triggers
in the \tnt code because it favors high ranked trigger compared to low ranked
(in case both apply, we prefer to use the most reliable, i.e., the less subject
to any interference such a BGP asymmetry). As indicated in
Listing~\ref{tnt.trig.check_trigger} (Sec.~\ref{tnt.trig}), we first check for
\rtla as it is more reliable than \frpla. DPR works better than BRPR, which is
obvious as it is triggered by \rtla (Juniper routers). For Invisible UHP, it is
worth noticing that the buddy bit, prior to BRPR or DPR revelation, was required
in nearly 25\% of the cases. In other cases, a simple BRPR or DPR revelation was
enough to get the tunnel content. UHP seems to be often filtered for a
particular FEC, e.g., only /32 host loopback addresses are advertised in LDP
with UHP while other FEC are advertised with PHP (BRPR) or are not injected at
all (DPR).

The column labeled ``mix'' corresponds to tunnels partially revealed thanks to
BRPR and partially with DPR. Typically, it comes from \textit{heterogeneous MPLS
clouds}. For instance, operators may deploy both Juniper and Cisco hardware
without any homogeneous prefixes distribution (i.e., local prefix for Juniper,
all prefixes for Cisco -- See Sec.~\ref{background.cp} for details). Note that
it is also possible that the UHP and PHP label popping techniques co-exist when
using our backward recursive path revelation (BRPR). Although not explained in
Sec.~\ref{tnt} for clarity reasons, \tnt can deal with those more complex
situations, making the tool quite robust to pitfalls encountered in the wild
Internet (5\% of the Invisible tunnels encountered).

Finally, the column labeled ``\unknown'' corresponds to one hop tunnels where
DPR and BRPR cannot be distinguished. This large proportion (20\%) of very short
Invisible tunnels is aligned with previous works that already noticed the
proportion of short Explicit tunnels~\cite{mpls-ccr,mpls-sommers,mpls-lpr}.


\section{Related Work}\label{related}
For years now, \traceroute has been used as the main tool for discovering the
Internet topology~\cite{survey}.  Multiple extensions have been provided to
circumvent \traceroute limits.

Doubletree~\cite{dt,dtBeverly} has been proposed for improving the cooperation
between scattered \traceroute vantage points, reducing so the probing
redundancy. Paris traceroute~\cite{parisTraceroute} has been developed for
fixing issues related to IP load balancing, avoiding so false links between IP
interfaces. \texttt{tracebox}~\cite{tracebox} extends \traceroute for revealing
the presence of middleboxes along a path.  YARRP~\cite{yarrp} provides
techniques for speeding up the \traceroute probing process.  Reverse
\traceroute~\cite{reverseTraceroute} is able to provide the reverse path (i.e.,
from the target back to the vantage point).  Passenger~\cite{sidecar} and
Discarte~\cite{discarte} extend \traceroute with the IP record route option.
Marchetta et al.~\cite{alternative} have proposed to use the ICMP Parameter
Problem in addition to Record Route option in \traceroute.  Finally,
\texttt{tracenet}~\cite{tracenet} mimics \traceroute for discovering
subnetworks.

\tnt is also in the scope of the \dfn{hidden router} issue, i.e., any device
that does not decrement the TTL causing the device to be transparent to
\traceroute probing.  Discarte and Passenger, through the use of IP Record Route
Option, allows, to some extent, to reveal hidden routers along a path.
\textsc{Drago}~\cite{drago} considers the ICMP Timestamp for also detecting
hidden routers.  \tnt goes beyond those solutions as it does not rely on ICMP
messages and IP option that are, generally, filtered by operators either locally
(i.e., the option/message is turned off on the router) or for transit packets
(i.e., edge routers do not forward those particular packets).\footnote{It has
been, however, demonstrated recently that IP Record Route option might still
find a suitable usage in Internet measurements if used with
prudence~\cite{recordRoute}.} \tnt only relies on standard messages
(\echorequest/\echoreply and \ttlexceeded) that are implemented and used by the
vast majority of routers and, as such, has the potential to reveal much more
information.

MPLS tunnels discovery has been the subject of several researches those last
years.  In particular, Sommers et al.~\cite{mpls-sommers} examined the
characteristics of MPLS deployments that are explicitly identified using RFC4950
extensions, as observed in CAIDA's topology data.  Donnet et al.~\cite{mpls-ccr}
proposed the first classification of MPLS tunnels according to the relationship
between MPLS and \traceroute.  This paper is a revision of Donnet et al.'s work
in light of a deeper understanding of MPLS mechanisms, in particular for hidden
tunnels (Opaque, Invisible PHP and UHP). More recently, Vanaubel et
al.~\cite{mpls-invisible} have proposed techniques for inferring and possibly
revealing hidden tunnels: \frpla, \rtla, \brpr, and \dpr. \frpla and \rtla were
initially not used as triggers for measurements (as we are doing in this paper
with \tnt by extending those techniques in many aspects) but rather as a way to
infer an hidden tunnel length. Vanaubel et al.~directed their measurements
towards pre-identified high degree routers with the ITDK dataset used as a
source for triggering specific measurements (as they were suspected to be the
exit point of a large number of hidden MPLS tunnels). As such, Vanaubel et al.
did not provide any integrated measurement tool, on the contrary to \tnt, in
which MPLS tunnels are discovered on the fly (\tnt does not rely on any kind of
external dataset).


\section{Conclusion}\label{ccl}
In this paper, we revise the MPLS classification proposed by Donnet et
al.~\cite{mpls-ccr} and introduce \tnt (\underline{\textbf{T}}race the
\underline{\textbf{N}}aughty \underline{\textbf{T}}unnels) that is an extension
to Paris traceroute for revealing MPLS tunnels along a path. As such, \tnt has
the potential to reveal more complete information on the exact Internet
topology. We provide accurate IP level tracing functions leading so to better
Internet models. For instance, it has been shown that Invisible tunnels have an
impact on Internet basic graph properties~\cite{mpls-invisible}). Our fully
integrated tool reveals, or at least detect, all kind of tunnels in two simple
stages: first, it uses indicators and triggers to respectively classify and
possibly tag tunnels as hidden, second it reveals the tagged tunnel content if
any. \tnt has the capacity to unveil the MPLS ecosystem deployed by operators.
Recent works on MPLS discovery have revealed that MPLS is largely deployed by
most ISP~\cite{mpls-ccr,mpls-lpr, mpls-sommers}. By running \tnt on a daily (or
nearly daily) basis from the Archipelago platform, we expect to see numerous
researches using our tool and data to mitigate the impact of MPLS on the
Internet topology. \tnt has been developed with a reproducibility perspective.
As such, it is freely available, as well as our dataset and scripts used for
processing data.\textsuperscript{\ref{footnote.web}}


\section*{Acknowledgments}
Authors would like to thank kc claffy and her team at \textsc{Caida} for letting
them deploying \tnt on the Archipelago infrastructure.  In addition, part of Mr.
Vanaubel's work was supported by an internship at \textsc{Caida}, under the
direction of Young Hyun.


\small{
\balance

}


\onecolumn
\begin{center}
\textbf{\huge{Appendix}}
\end{center}

\setcounter{section}{0}

This appendix illustrates the validation of \tnt through GNS-3 emulations.
Multiple configurations have been tested (and even more are proposed on the
website\textsuperscript{\ref{footnote.web}} and can be setup using the scripts
and the data online). Note that we use the version 2.1.5 of GNS3 to export the
so-called portable configurations. \tnt is able to deal with all those
configurations (both in the wild and with the ones emulated in GNS3), making it
a pretty robust tool. However, in this report we use another version of our tool
to simplify the output. The output of \tnt slightly differ but the conclusions
are the same.

\begin{figure*}[!ht]
  \begin{center}
    \begin{subfigure}[b]{16cm}
      \includegraphics[width=16cm]{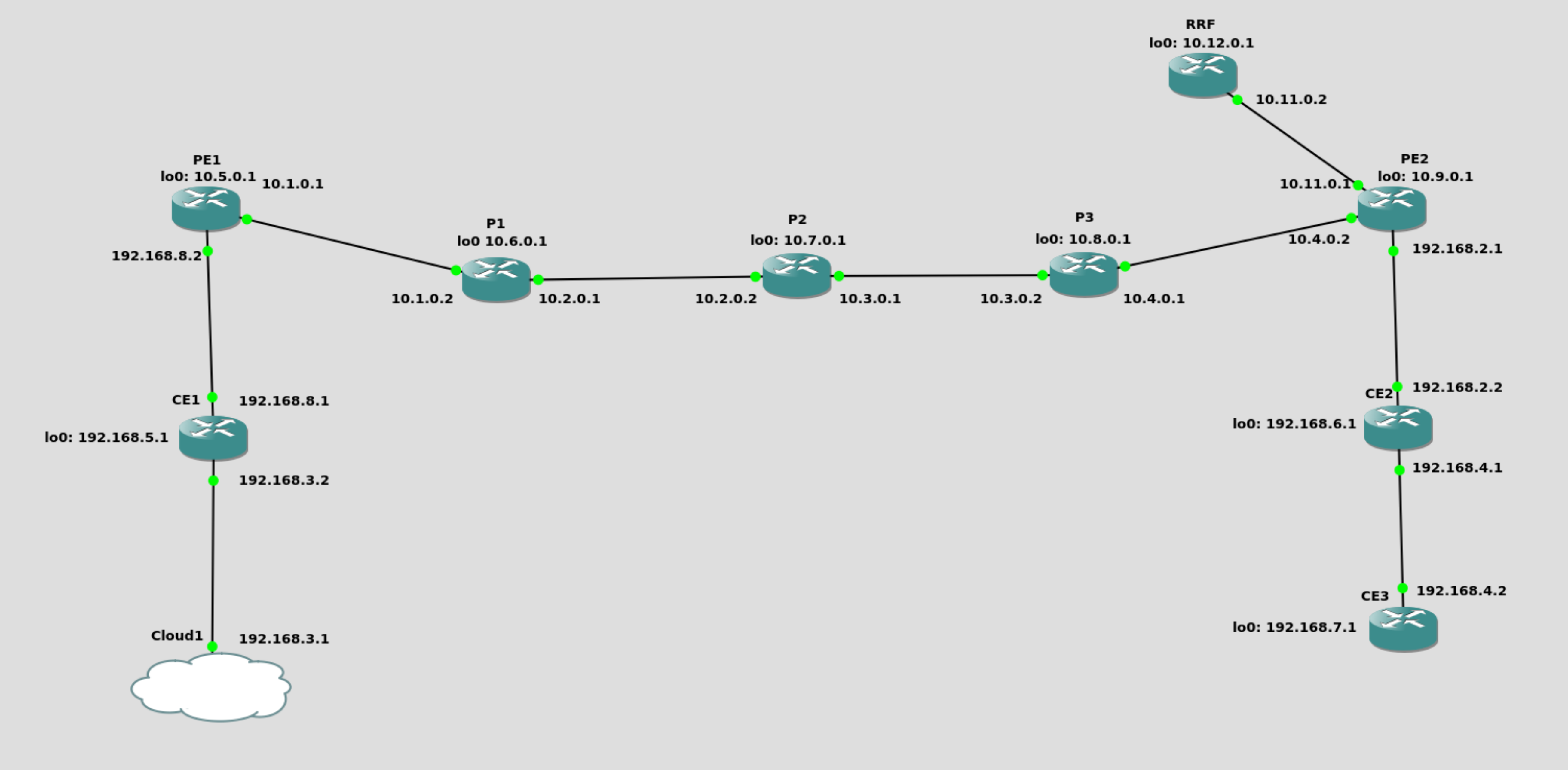}
      \caption{Cisco topology.  PE1 is the Ingress LER, PE2 the
      Egress LER, the LSP is set up between PE1 and the EH (P3 or PE2).  The \tnt target (i.e.,
      the argument of \texttt{trace\_naughty\_tunnel()} function -- See
      Listing~\ref{tnt.overview.trace_tunnel}) is the loopback address of CE3.}
      \label{appendix.topo.cisco}
    \end{subfigure}

\medskip
\medskip

    \begin{subfigure}[b]{16cm}
      \includegraphics[width=16cm]{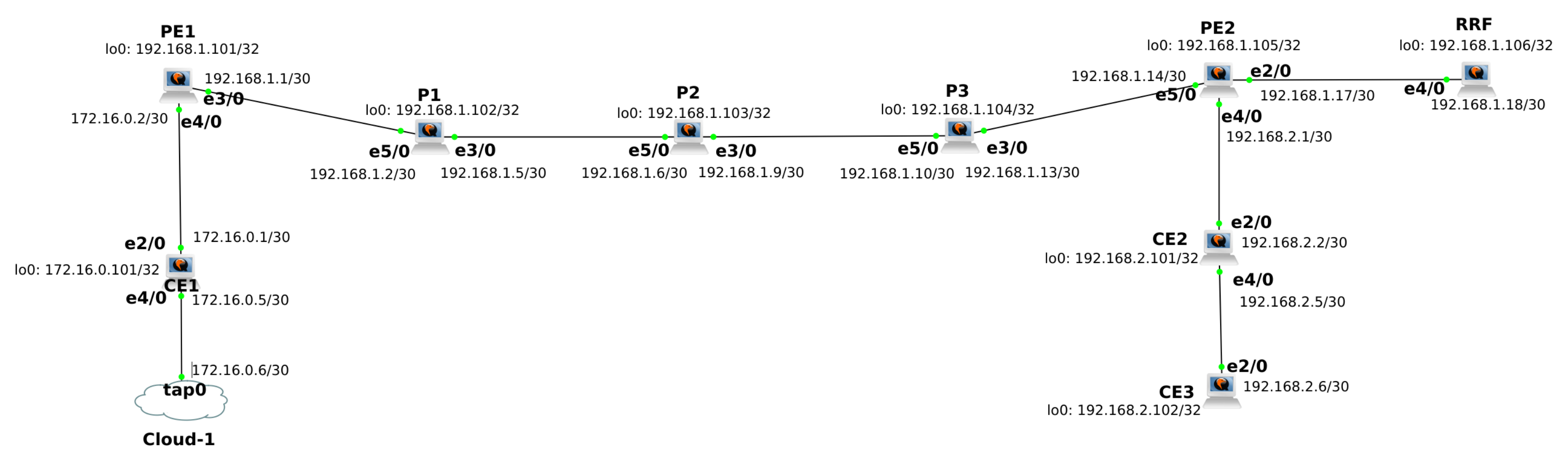}
      \caption{Juniper topology.  PE1 is the Ingress LER, PE2 the
      Egress LER, the LSP is set up between PE1 and the EH (P3 or PE2).  The \tnt target (i.e.,
      the argument of \texttt{trace\_naughty\_tunnel()} function -- See
      Listing~\ref{tnt.overview.trace_tunnel}) is the loopback address of CE3.}
      \label{appendix.topo.juniper}
    \end{subfigure}
  \end{center}
  \caption{Topology used for GNS-3 tests}
  \label{appenix.topo}
\end{figure*}

\pagebreak

\section{P2P Circuits}\label{appendix.p2p}

\subsection{Explicit Tunnels Validation}\label{appendix.explicit}
We first review Explicit tunnels, i.e., tunnels with RFC4950 and \tpropagate
enabled (see Sec.~\ref{taxo}).

In the following, we distinguish Cisco (Appendix~\ref{appendix.explicit.cisco})
and Juniper IP topologies (Appendix~\ref{appendix.explicit.juniper}) and
configurations. In particular, with Cisco configurations, PHP (LSE popped by P3)
is distinguished from UHP (LSE popped by Egress LER).

For each case, we provide the configuration of routers as well as the simplified \tnt
output. Indicators and triggers (see Sec.~\ref{tnt.trig}) are provided, as well
as raw ICMP \ttlexceeded and ICMP \echoreply TTLs.

\subsubsection{Cisco Explicit Configurations}\label{appendix.explicit.cisco}
All configurations presented here were run on the IP topology provided by
Fig.~\ref{appendix.topo.cisco}.

The first example provides an Explicit tunnel deployed with PHP, under Cisco IOS 15.2.
The \tnt behavior is the one expected.

\begin{cisco}[title=IOS 15.2 -- Explicit PHP]
PE1
version 15.2
mpls label protocol ldp
router bgp 3333
 redistribute connected
 redistribute ospf 10
 neighbor 10.12.0.1 remote-as 3333
 neighbor 10.12.0.1 next-hop-self
 neighbor 192.168.8.1 remote-as 1024
 neighbor 192.168.8.1 next-hop-self

PE2
version 15.2
mpls label protocol ldp
router bgp 3333
 redistribute connected
 redistribute ospf 10
 neighbor 10.12.0.1 remote-as 3333
 neighbor 10.12.0.1 next-hop-self
 neighbor 192.168.2.2 remote-as 2048
 neighbor 192.168.2.2 next-hop-self

P1
version 15.2
mpls label protocol ldp
router bgp 3333
 neighbor 10.12.0.1 remote-as 3333

P2
version 15.2
mpls label protocol ldp
router bgp 3333
 neighbor 10.12.0.1 remote-as 3333

P3
version 15.2
mpls label protocol ldp
router bgp 3333
 neighbor 10.12.0.1 remote-as 3333
\end{cisco}

\begin{cisco}[title=\tnt running over IOS 15.2 -- Explicit PHP]
Launching TNT: 192.168.7.1 (192.168.7.1)

  1  left.CE1 (192.168.3.2)  <255,255> [frpla = 0][qttl = 1][uturn = 0] 27.083 ms
  2  left.PE1 (192.168.8.2)  <254,254> [frpla = 0][qttl = 1][uturn = 0] 19.895 ms
  3  left.P1 (10.1.0.2)  <247,253> [frpla = 6][qttl = 1][uturn = 6][MPLS LSE | Label : 19 | LSE-TTL : 1]  80.598 ms
  4  left.P2 (10.2.0.2)  <248,252> [frpla = 4][qttl = 2][uturn = 4][MPLS LSE | Label : 20 | LSE-TTL : 1]  69.875 ms
  5  left.P3 (10.3.0.2)  <251,251> [frpla = 0][qttl = 1][uturn = 0][MPLS LSE | Label : 20 | LSE-TTL : 1]  68.98 ms
  6  left.PE2 (10.4.0.2)  <250,250> [frpla = 0][qttl = 1][uturn = 0] 78.17 ms
  7  left.CE2 (192.168.2.2)  <249,249> [frpla = 0][qttl = 1][uturn = 0] 78.957 ms
  8  192.168.4.2 (192.168.4.2)  <248,248> [frpla = 0][qttl = 1][uturn = 0] 110.598 ms
\end{cisco}

The next two configurations illustrate UHP with both IOS 12.4 and IOS 15.2.
\tnt works as expected and shows two examples of MPLS TTL processing specifically with UHP.
With the 12.4 IOS, we see the null label while it is hidden with the 15.2 IOS.
In addition, we can see that UHP tunnels show a \uturn signature different from PHP tunnels.
This difference results from the way time-exceeded messages are handled by the LSRs. In both cases,
the time-exceeded message is forwarded to the EH which replies using its own IP forwarding table.
The EH changes depending on the configuration: P3 for PHP (here the EH is the PH), and PE2 for UHP
(here the EH is the Egress LER). Indeed, we can see that the \uturn difference disappears at the respective EH.

\begin{cisco}[title=IOS 12.4 -- Explicit UHP]
PE1
version 12.4
mpls label protocol ldp
mpls ldp explicit-null
router bgp 3333
 redistribute connected
 redistribute ospf 10
 neighbor 10.12.0.1 remote-as 3333
 neighbor 10.12.0.1 next-hop-self
 neighbor 192.168.8.1 remote-as 1024
 neighbor 192.168.8.1 next-hop-self

PE2
version 12.4
mpls label protocol ldp
mpls ldp explicit-null
router bgp 3333
 redistribute connected
 redistribute ospf 10
 neighbor 10.12.0.1 remote-as 3333
 neighbor 10.12.0.1 next-hop-self
 neighbor 192.168.2.2 remote-as 2048
 neighbor 192.168.2.2 next-hop-self

 P1
 version 12.4
 mpls label protocol ldp
 mpls ldp explicit-null
 router bgp 3333
  neighbor 10.12.0.1 remote-as 3333

 P2
 version 12.4
 mpls label protocol ldp
 mpls ldp explicit-null
 router bgp 3333
  neighbor 10.12.0.1 remote-as 3333

 P3
 version 12.4
 mpls label protocol ldp
 mpls ldp explicit-null
 router bgp 3333
  neighbor 10.12.0.1 remote-as 3333
\end{cisco}

\begin{cisco}[title=\tnt running over IOS 12.4 -- Explicit UHP]
Launching TNT: 192.168.7.1 (192.168.7.1)

  1  left.CE1 (192.168.3.2)  <255,255> [frpla = 0][qttl = 1][uturn = 0] 22.651 ms
  2  192.168.8.2 (192.168.8.2)  <254,254> [frpla = 0][qttl = 1][uturn = 0] 230.326 ms
  3  left.P1 (10.1.0.2)  <247,253> [frpla = 6][qttl = 1][uturn = 6][MPLS LSE | Label : 22 | LSE-TTL : 1]  263.686 ms
  4  left.P2 (10.2.0.2)  <248,252> [frpla = 4][qttl = 2][uturn = 4][MPLS LSE | Label : 22 | LSE-TTL : 1]  358.238 ms
  5  left.P3 (10.3.0.2)  <249,251> [frpla = 2][qttl = 3][uturn = 2][MPLS LSE | Label : 16 | LSE-TTL : 1]  374.214 ms
  6  left.PE2 (10.4.0.2)  <250,250> [frpla = 0][qttl = 1][uturn = 0][MPLS LSE | Label : 0 | LSE-TTL : 1]  418.696 ms
  7  left.CE2 (192.168.2.2)  <249,249> [frpla = 0][qttl = 1][uturn = 0] 655.848 ms
  8  192.168.4.2 (192.168.4.2)  <248,248> [frpla = 0][qttl = 1][uturn = 0] 513.054 ms
\end{cisco}

\begin{cisco}[title=IOS 15.2 -- Explicit UHP]
PE1
version 15.2
mpls label protocol ldp
mpls ldp explicit-null
router bgp 3333
 redistribute connected
 redistribute ospf 10
 neighbor 10.12.0.1 remote-as 3333
 neighbor 10.12.0.1 next-hop-self
 neighbor 192.168.8.1 remote-as 1024
 neighbor 192.168.8.1 next-hop-self

PE2
version 15.2
mpls label protocol ldp
mpls ldp explicit-null
router bgp 3333
 redistribute connected
 redistribute ospf 10
 neighbor 10.12.0.1 remote-as 3333
 neighbor 10.12.0.1 next-hop-self
 neighbor 192.168.2.2 remote-as 2048
 neighbor 192.168.2.2 next-hop-self

P1
version 15.2
mpls label protocol ldp
mpls ldp explicit-null
router bgp 3333
 neighbor 10.12.0.1 remote-as 3333

P2
version 15.2
mpls label protocol ldp
mpls ldp explicit-null
router bgp 3333
 neighbor 10.12.0.1 remote-as 3333

P3
version 15.2
mpls label protocol ldp
mpls ldp explicit-null
router bgp 3333
 neighbor 10.12.0.1 remote-as 3333
\end{cisco}

\begin{cisco}[title=\tnt running over IOS 15.2 -- Explicit UHP]
Launching TNT: 192.168.7.1 (192.168.7.1)

  1  left.CE1 (192.168.3.2)  <255,255> [frpla = 0][qttl = 1][uturn = 0] 7.64 ms
  2  left.PE1 (192.168.8.2)  <254,254> [frpla = 0][qttl = 1][uturn = 0] 39.87 ms
  3  left.P1 (10.1.0.2)  <247,253> [frpla = 6][qttl = 1][uturn = 6][MPLS LSE | Label : 19 | LSE-TTL : 1]  100.632 ms
  4  left.P2 (10.2.0.2)  <248,252> [frpla = 4][qttl = 2][uturn = 4][MPLS LSE | Label : 20 | LSE-TTL : 1]  80.453 ms
  5  left.P3 (10.3.0.2)  <249,251> [frpla = 2][qttl = 3][uturn = 2][MPLS LSE | Label : 20 | LSE-TTL : 1]  100.815 ms
  6  left.PE2 (10.4.0.2)  <250,250> [frpla = 0][qttl = 1][uturn = 0] 109.089 ms
  7  left.CE2 (192.168.2.2)  <249,249> [frpla = 0][qttl = 1][uturn = 0] 98.817 ms
  8  192.168.4.2 (192.168.4.2)  <248,248> [frpla = 0][qttl = 1][uturn = 0] 119.842 ms
\end{cisco}

\subsubsection{Juniper Explicit Configurations}\label{appendix.explicit.juniper}
All configurations presented here were run on the topology provided by
Fig.~\ref{appendix.topo.juniper}.

For Explicit tunnels, Juniper Olive and VMX behave the same.  We first provide
the configuration and \tnt output for Explicit tunnels without \uturn effect.

\begin{cisco}[title=VMX -- Explicit PHP (default configuration)]
PE1
propagate ttl

PE2
propagate ttl

P1
propagate ttl

P2
propagate ttl

P3
propagate ttl
\end{cisco}

\begin{cisco}[title=\tnt running over VMX - Explicit PHP (default configuration)]
Launching TNT: 192.168.2.102 (192.168.2.102)

  1  CE1 ( 172.16.0.5)  <255,64> [frpla = 0][qttl = 1][uturn = 0] 2.682 ms
  2  PE1 ( 172.16.0.2)  <254,63> [frpla = 0][qttl = 1][uturn = 0] 4.603 ms
  3  left.P1 (192.168.1.2)  <253,62> [frpla = 0][qttl = 1][uturn = 0][MPLS LSE | Label : 299824 | LSE-TTL : 1]  6.362 ms
  4  left.P2 (192.168.1.6)  <252,61> [frpla = 0][qttl = 1][uturn = 0][MPLS LSE | Label : 299792 | LSE-TTL : 1]  8.451 ms
  5  left.P3 (192.168.1.10)  <251,60> [frpla = 0][qttl = 1][uturn = 0][MPLS LSE | Label : 299792 | LSE-TTL : 1]  8.557 ms
  6  left.PE2 (192.168.1.14)  <250,59> [frpla = 0][qttl = 1][uturn = 0] 8.285 ms
  7  CE2 (192.168.2.2)  <249,58> [frpla = 0][qttl = 1][uturn = 0] 8.09 ms
  8  CE3 (192.168.2.102)  <248,57> [frpla = 0][qttl = 1][uturn = 0] 8.142 ms
\end{cisco}

On the contrary to Cisco configuration, Juniper does not exhibit the \uturn effect.
When the LSE-TTL of a packet expires, the LSR does not send the ICMP
\ttlexceeded to the EH which then forwards the packets on its own to the
probing source, it replies the same with respect to other probes (e.g.,
\echorequest) using its own IP forwarding table if available -- resulting in
general in a shorter return path (see Sec.~\ref{tnt.trig}). The configuration must be
explicitly stated with the \texttt{icmp-tunneling} as provided below.

\begin{cisco}[title=VMX -- Explicit PHP (\texttt{icmp-tunneling} configuration)]
PE1
propagate ttl
icmp-tunneling

PE2
propagate ttl
icmp-tunneling

P1
propagate ttl
icmp-tunneling

P2
propagate ttl
icmp-tunneling

P3
propagate ttl
icmp-tunneling
\end{cisco}

\begin{cisco}[title=\tnt running over VMX -- Explicit PHP (\texttt{icmp-tunneling} configuration)]
Launching TNT: 192.168.2.102 (192.168.2.102)

  1  CE1 ( 172.16.0.5)  <255,64> [frpla = 0][qttl = 1][uturn = 0] 2.034 ms
  2  PE1 ( 172.16.0.2)  <254,63> [frpla = 0][qttl = 1][uturn = 0] 4.646 ms
  3  left.P1 (192.168.1.2)  <246,62> [frpla = 7][qttl = 1][uturn = 7][MPLS LSE | Label : 299824 | LSE-TTL : 1]  11.424 ms
  4  left.P2 (192.168.1.6)  <247,61> [frpla = 5][qttl = 1][uturn = 5][MPLS LSE | Label : 299824 | LSE-TTL : 1]  7.994 ms
  5  left.P3 (192.168.1.10)  <251,60> [frpla = 0][qttl = 1][uturn = 0][MPLS LSE | Label : 299824 | LSE-TTL : 1]  6.252 ms
  6  left.PE2 (192.168.1.14)  <250,59> [frpla = 0][qttl = 1][uturn = 0] 8.585 ms
  7  CE2 (192.168.2.2)  <249,58> [frpla = 0][qttl = 1][uturn = 0] 9.369 ms
  8  CE3 (192.168.2.102)  <248,57> [frpla = 0][qttl = 1][uturn = 0] 9.232 ms
\end{cisco}

\subsection{Opaque Tunnels Validation (Cisco only)}\label{appendix.opaque}
Opaque tunnels only occur with Cisco routers, in some particular configuration
(see Sec.~\ref{taxo} for details). The topology used for GNS-3
emulation is the one provided by Fig.~\ref{appendix.topo.cisco}.  We only show
tests for IOS 15.2 as the situation is the same with IOS 12.4. In our example,
we were able to reveal the content of the Opaque tunnel through BRPR, on the
contrary to in the wild \tnt deployment where Opaque tunnels revelation did not
work that much (see Sec.~\ref{tnt_quantif}).  We see thus here a difference
between theory and practice.

\begin{cisco}[title=IOS 15.2 -- Opaque PHP]
PE1
version 15.2
mpls label protocol ldp
no  propagate-ttl
router bgp 3333
 redistribute connected
 redistribute ospf 10
 neighbor 10.12.0.1 remote-as 3333
 neighbor 192.168.8.1 remote-as 1024

PE2
version 15.2
mpls label protocol ldp
no  propagate-ttl
router bgp 3333
 redistribute connected
 redistribute ospf 10
 neighbor 10.12.0.1 remote-as 3333
 neighbor 192.168.6.1 remote-as 2048
 neighbor 192.168.6.1 ebgp-multihop 2

P1
version 15.2
mpls label protocol ldp
no  propagate-ttl
router bgp 3333
 neighbor 10.12.0.1 remote-as 3333

P2
version 15.2
mpls label protocol ldp
no  propagate-ttl
router bgp 3333
 neighbor 10.12.0.1 remote-as 3333

P3
version 15.2
mpls label protocol ldp
no  propagate-ttl
router bgp 3333
 neighbor 10.12.0.1 remote-as 3333
\end{cisco}

\begin{cisco}[title=\tnt running over IOS 15.2 -- Opaque PHP]
Launching TNT: 192.168.7.1 (192.168.7.1)

  1  left.CE1 (192.168.3.2)  <255,255> [frpla = 0][qttl = 1][uturn = 0] 25.164 ms
  2  left.PE1 (192.168.8.2)  <254,254> [frpla = 0][qttl = 1][uturn = 0] 40.06 ms

    OPAQUE | Length estimation : 3 | Revealed : 3 (difference : 0)
     2.1 [REVEALED] left.P1 (10.1.0.2)  <253,253> [frpla = 0][qttl = 1][uturn = 0] 40.008 ms  - step 2
     2.2 [REVEALED] left.P2 (10.2.0.2)  <252,252> [frpla = 0][qttl = 1][uturn = 0] 40.058 ms  - step 1
     2.3 [REVEALED] left.P3 (10.3.0.2)  <251,251> [frpla = 0][qttl = 1][uturn = 0] 90.301 ms  - step 0

  3  left.PE2 (10.4.0.2)  <250,250> [frpla = 3][qttl = 1][uturn = 0][MPLS LSE | Label : 16 | LSE-TTL : 252]  110.408 ms
  4  left.CE2 (192.168.2.2)  <250,250> [frpla = 2][qttl = 1][uturn = 0] 80.195 ms
  5  192.168.4.2 (192.168.4.2)  <250,250> [frpla = 1][qttl = 1][uturn = 0] 132.331 ms
\end{cisco}

\subsection{Invisible Tunnels Validation}\label{appendix.invisible}
This section discusses Invisible tunnels, i.e., tunnels with
the \notpropagate option enabled (see Sec.~\ref{taxo}).

We do a distinction between Cisco
(Appendix~\ref{appendix.invisible.cisco}) and Juniper configurations
(Appendix~\ref{appendix.invisible.juniper}).  PHP (LSE popped by P3) is also
distinguished from UHP (LSE popped by Egress LER).

For each case, we provide the configuration of routers as well as the \tnt
output.  Indicators and triggers (see Sec.~\ref{tnt.trig}) are provided, as well
as ICMP \ttlexceeded and ICMP \echoreply TTLs.

\subsubsection{Invisible Cisco Configurations}\label{appendix.invisible.cisco}
All configurations presented here were run on the topology provided by
Fig.~\ref{appendix.topo.cisco}.

\begin{cisco}[title=IOS 15.2 -- Invisible PHP]
PE1
version 15.2
mpls label protocol ldp
no propagate-ttl
router bgp 3333
 redistribute connected
 redistribute ospf 10
 neighbor 10.12.0.1 remote-as 3333
 neighbor 10.12.0.1 next-hop-self
 neighbor 192.168.8.1 remote-as 1024
 neighbor 192.168.8.1 next-hop-self

PE2
version 15.2
mpls label protocol ldp
no  propagate-ttl
router bgp 3333
 redistribute connected
 redistribute ospf 10
 neighbor 10.12.0.1 remote-as 3333
 neighbor 10.12.0.1 next-hop-self
 neighbor 192.168.2.2 remote-as 2048
 neighbor 192.168.2.2 next-hop-self

P1
version 15.2
mpls label protocol ldp
no  propagate-ttl
router bgp 3333
 neighbor 10.12.0.1 remote-as 3333

P2
version 15.2
mpls label protocol ldp
no  propagate-ttl
router bgp 3333
 neighbor 10.12.0.1 remote-as 3333

P3
version 15.2
mpls label protocol ldp
no  propagate-ttl
router bgp 3333
 neighbor 10.12.0.1 remote-as 3333
\end{cisco}

\begin{cisco}[title=\tnt running over IOS 15.2 -- Invisible PHP]
Launching TNT: 192.168.7.1 (192.168.7.1)

  1  left.CE1 (192.168.3.2)  <255,255> [frpla = 0][qttl = 1][uturn = 0] 7.52 ms
  2  left.PE1 (192.168.8.2)  <254,254> [frpla = 0][qttl = 1][uturn = 0] 29.927 ms

    FRPLA | Length estimation : 3 | Revealed : 3 (difference : 0)
     2.1 [REVEALED] left.P1 (10.1.0.2)  <253,253> [frpla = 0][qttl = 1][uturn = 0] 50.051 ms  - step 2
     2.2 [REVEALED] left.P2 (10.2.0.2)  <252,252> [frpla = 0][qttl = 1][uturn = 0] 60.102 ms  - step 1
     2.3 [REVEALED] left.P3 (10.3.0.2)  <251,251> [frpla = 0][qttl = 1][uturn = 0] 59.876 ms  - step 0

  3  left.PE2 (10.4.0.2)  <250,250> [frpla = 3][qttl = 1][uturn = 0] 80.38 ms
  4  left.CE2 (192.168.2.2)  <250,250> [frpla = 2][qttl = 1][uturn = 0] 69.89 ms
  5  192.168.4.2 (192.168.4.2)  <250,250> [frpla = 1][qttl = 1][uturn = 0] 99.833 ms
\end{cisco}

The configuration for running standard Cisco Invisible UHP tunnels is provided
below. Such a configuration might be revealed through BRPR thanks to the \dupip
trigger.

\begin{cisco}[title=IOS 15.2 -- Invisible UHP]
PE1
version 15.2
mpls label protocol ldp
no  propagate-ttl
mpls ldp explicit-null
router bgp 3333
 redistribute connected
 redistribute ospf 10
 neighbor 10.12.0.1 remote-as 3333
 neighbor 10.12.0.1 next-hop-self
 neighbor 192.168.8.1 remote-as 1024
 neighbor 192.168.8.1 next-hop-self

PE2
version 15.2
mpls label protocol ldp
no  propagate-ttl
mpls ldp explicit-null
router bgp 3333
 redistribute connected
 redistribute ospf 10
 neighbor 10.12.0.1 remote-as 3333
 neighbor 10.12.0.1 next-hop-self
 neighbor 192.168.2.2 remote-as 2048
 neighbor 192.168.2.2 next-hop-self

P1
version 15.2
mpls label protocol ldp
no  propagate-ttl
mpls ldp explicit-null
router bgp 3333
 neighbor 10.12.0.1 remote-as 3333

P2
version 15.2
mpls label protocol ldp
no  propagate-ttl
mpls ldp explicit-null
router bgp 3333
 neighbor 10.12.0.1 remote-as 3333

P3
version 15.2
mpls label protocol ldp
no  propagate-ttl
mpls ldp explicit-null
router bgp 3333
 neighbor 10.12.0.1 remote-as 3333
\end{cisco}

\begin{cisco}[title=\tnt running over  IOS 15.2 -- Invisible UHP]
Launching TNT: 192.168.7.1 (192.168.7.1)

  1  left.CE1 (192.168.3.2)  <255,255> [frpla = 0][qttl = 1][uturn = 0] 3.157 ms
  2  left.PE1 (192.168.8.2)  <254,254> [frpla = 0][qttl = 1][uturn = 0] 29.92 ms

    Duplicate IP (Egress : 192.168.2.2) | Length estimation : 1 | Revealed : 4 (difference : 3)
     2.1 [REVEALED] left.P1 (10.1.0.2)  <253,253> [frpla = 0][qttl = 1][uturn = 0] 50.043 ms  - step 4  (Buddy used)
     2.2 [REVEALED] left.P2 (10.2.0.2)  <253,253> [frpla = 0][qttl = 1][uturn = 0] 49.778 ms  - step 3  (Buddy used)
     2.3 [REVEALED] left.P3 (10.3.0.2)  <253,253> [frpla = 0][qttl = 1][uturn = 0] 69.834 ms  - step 2  (Buddy used)
     2.4 [REVEALED] left.PE2 (10.4.0.2)  <253,253> [frpla = 0][qttl = 1][uturn = 0] 80.594 ms  - step 1  (Buddy used)

  3  left.CE2 (192.168.2.2)  <252,252> [frpla = 1][qttl = 1][uturn = 0] 80.08 ms
  4  left.CE2 (192.168.2.2)  <252,252> [frpla = 0][qttl = 1][uturn = 0] 89.891 ms
  5  192.168.4.2 (192.168.4.2)  <251,251> [frpla = 0][qttl = 1][uturn = 0] 107.579 ms
\end{cisco}

With Cisco routers, it is possible to mimic an Invisible UHP tunnel with a
Juniper per loopback configuration (i.e., by filtering addresses to /32 border
prefixes), meaning that the tunnel content might be revealed through DPR, thanks
to the \dupip trigger.  Such a configuration is achieved with the
\texttt{allocate global host-routes} command.

\begin{cisco}[title=IOS 15.2 -- Invisible UHP (\texttt{allocate global host route} configuration)]
PE1
version 15.2
mpls label protocol ldp
no  propagate-ttl
mpls ldp explicit-null
mpls ldp label
 allocate global host-routes
router bgp 3333
 redistribute connected
 redistribute ospf 10
 neighbor 10.12.0.1 remote-as 3333
 neighbor 10.12.0.1 next-hop-self
 neighbor 192.168.8.1 remote-as 1024
 neighbor 192.168.8.1 next-hop-self

PE2
version 15.2
mpls label protocol ldp
no  propagate-ttl
mpls ldp explicit-null
mpls ldp label
 allocate global host-routes
router bgp 3333
 redistribute connected
 redistribute ospf 10
 neighbor 10.12.0.1 remote-as 3333
 neighbor 10.12.0.1 next-hop-self
 neighbor 192.168.2.2 remote-as 2048
 neighbor 192.168.2.2 next-hop-self

P1
version 15.2
mpls label protocol ldp
no  propagate-ttl
mpls ldp explicit-null
mpls ldp label
 allocate global host-routes
router bgp 3333
 neighbor 10.12.0.1 remote-as 3333

P2
version 15.2
mpls label protocol ldp
no  propagate-ttl
mpls ldp explicit-null
mpls ldp label
 allocate global host-routes
router bgp 3333
 neighbor 10.12.0.1 remote-as 3333

P3
version 15.2
mpls label protocol ldp
no  propagate-ttl
mpls ldp explicit-null
mpls ldp label
 allocate global host-routes
router bgp 3333
 neighbor 10.12.0.1 remote-as 3333
\end{cisco}

\begin{cisco}[title=\tnt running over IOS 15.2 -- Invisible UHP (\texttt{allocate global host route} configuration)]
Launching TNT: 192.168.7.1 (192.168.7.1)

  1  left.CE1 (192.168.3.2)  <255,255> [frpla = 0][qttl = 1][uturn = 0] 8.091 ms
  2  left.PE1 (192.168.8.2)  <254,254> [frpla = 0][qttl = 1][uturn = 0] 39.867 ms

    Duplicate IP (Egress : 10.1.0.2) | Length estimation : 1 | Revealed : 4 (difference : 3)
     2.1 [REVEALED] left.P1 (10.1.0.2)  <253,253> [frpla = 0][qttl = 1][uturn = 0] 39.788 ms  - step 2
     2.2 [REVEALED] left.P2 (10.2.0.2)  <253,253> [frpla = 0][qttl = 1][uturn = 0] 49.573 ms  - step 2
     2.3 [REVEALED] left.P3 (10.3.0.2)  <253,253> [frpla = 0][qttl = 1][uturn = 0] 70.094 ms  - step 2
     2.4 [REVEALED] left.PE2 (10.4.0.2)  <253,253> [frpla = 0][qttl = 1][uturn = 0] 89.171 ms  - step 1  ( Buddy used )

  3  left.CE2 (192.168.2.2)  <252,252> [frpla = 1][qttl = 1][uturn = 0] 120.546 ms
  4  left.CE2 (192.168.2.2)  <252,252> [frpla = 0][qttl = 1][uturn = 0] 89.892 ms
  5  192.168.4.2 (192.168.4.2)  <251,251> [frpla = 0][qttl = 1][uturn = 0] 117.301 ms
\end{cisco}

It is also possible to build Invisible UHP tunnel in which the buddy mechanism
is not necessary (as we discover in the wild). Simply running BRPR will make the
tunnel content visible.  This configuration might be achieved with the
\texttt{ip access-list} command to enable Ultimate Hop Popping for external destinations only:

\begin{cisco}[title=IOS 15.2 -- Invisible UHP (\texttt{mpls ldp explicit-null [for prefix-acl]} configuration)]
PE1
version 15.2
mpls label protocol ldp
no  propagate-ttl
router bgp 3333
 redistribute connected
 redistribute ospf 10
 neighbor 10.12.0.1 remote-as 3333
 neighbor 10.12.0.1 next-hop-self
 neighbor 192.168.8.1 remote-as 1024
 neighbor 192.168.8.1 next-hop-self

PE2
version 15.2
mpls label protocol ldp
no  propagate-ttl
mpls ldp explicit-null for BRPR-wo-buddy
router bgp 3333
 redistribute connected
 redistribute ospf 10
 neighbor 10.12.0.1 remote-as 3333
 neighbor 10.12.0.1 next-hop-self
 neighbor 192.168.2.2 remote-as 2048
 neighbor 192.168.2.2 next-hop-self
ip access-list standard BRPR-wo-buddy
 permit 10.9.0.1
 deny any

P1
version 15.2
mpls label protocol ldp
no  propagate-ttl
router bgp 3333
 neighbor 10.12.0.1 remote-as 3333

P2
version 15.2
mpls label protocol ldp
no  propagate-ttl
router bgp 3333
 neighbor 10.12.0.1 remote-as 3333

P3
version 15.2
mpls label protocol ldp
no  propagate-ttl
router bgp 3333
 neighbor 10.12.0.1 remote-as 3333
\end{cisco}

\begin{cisco}[title=\tnt running over IOS 15.2 -- Invisible UHP (\texttt{mpls ldp explicit-null [for prefix-acl]} configuration)]
Launching TNT: 192.168.7.1 (192.168.7.1)

  1  192.168.3.2 (192.168.3.2)  <255,255> [frpla = 0][qttl = 1][uturn = 0] 7.299 ms
  2  192.168.8.2 (192.168.8.2)  <254,254> [frpla = 0][qttl = 1][uturn = 0] 14.921 ms

    Duplicate IP (Egress : 10.4.0.2) | Length estimation : 3 | Revealed : 4 (difference : 1)
     2.1 [REVEALED] 10.1.0.2 (10.1.0.2)  <253,253> [frpla = 0][qttl = 1][uturn = 0] 36.443 ms  - step 3
     2.2 [REVEALED] 10.2.0.2 (10.2.0.2)  <252,252> [frpla = 0][qttl = 1][uturn = 0] 35.879 ms  - step 2
     2.3 [REVEALED] 10.3.0.2 (10.3.0.2)  <251,251> [frpla = 0][qttl = 1][uturn = 0] 66.288 ms  - step 1
     2.4 [REVEALED] 10.4.0.2 (10.4.0.2)  <250,250> [frpla = 0][qttl = 1][uturn = 0] 64.19 ms  - step 0

  3  CE2 (192.168.2.2)  <250,250> [frpla = 3][qttl = 1][uturn = 0] 116.643 ms
  4  CE2 (192.168.2.2)  <250,250> [frpla = 2][qttl = 1][uturn = 0] 99.93 ms
  5  192.168.4.2 (192.168.4.2)  <250,250> [frpla = 1][qttl = 1][uturn = 0] 94.185 ms
\end{cisco}

\subsubsection{Juniper Invisible Configurations}\label{appendix.invisible.juniper}
All configurations presented here were run on the topology provided by
Fig.~\ref{appendix.topo.juniper}.

Juniper, with Olive OS, does not apply the \minttl{IP-TTL}{LSE-TTL} at the exit
of the MPLS cloud. As such, the \frpla trigger does not provide the return
tunnel length but is equal to 1 because the ingress LER process the incoming IP
TTL in a distinct way with respect to the origin of the packet (locally generated or
not).  Invisible PHP tunnel can, then, be revealed through DPR.  Juniper LSR
can be configured as followed:

\begin{cisco}[title=JunOS Olive -- Invisible PHP]
PE1
no-propagate-ttl

PE2
no-propagate-ttl

P1
no-propagate-ttl

P2
no-propagate-ttl

P3
no-propagate-ttl
\end{cisco}

\begin{cisco}[title=\tnt running over JunOS Olive -- Invisible PHP]
Launching TNT: 192.168.2.102 (192.168.2.102)

  1  CE1 ( 172.16.0.5)  <255,64> [frpla = 0][qttl = 1][uturn = 0] 0.638 ms
  2  PE1 ( 172.16.0.2)  <254,63> [frpla = 0][qttl = 1][uturn = 0] 1.898 ms

    FRPLA | Length estimation : 1 | Revealed : 3 (difference : 2)
     2.1 [REVEALED] left.P1 (192.168.1.2)  <253,62> [frpla = 0][qttl = 1][uturn = 0] 3.039 ms  - step 0
     2.2 [REVEALED] left.P2 (192.168.1.6)  <252,61> [frpla = 0][qttl = 1][uturn = 0] 3.951 ms  - step 0
     2.3 [REVEALED] left.P3 (192.168.1.10)  <252,61> [frpla = 0][qttl = 1][uturn = 0] 4.906 ms  - step 0

  3  left.PE2 (192.168.1.14)  <252,61> [frpla = 1][qttl = 1][uturn = 0] 7.043 ms
  4  CE2 (192.168.2.2)  <252,61> [frpla = 0][qttl = 1][uturn = 0] 6.891 ms
  5  CE3 (192.168.2.102)  <251,60> [frpla = 0][qttl = 1][uturn = 0] 8.978 ms
\end{cisco}

On the contrary to Olive, VMX applies the \minttl{IP-TTL}{LSE-TTL} function.  As
such, the behavior observed is the theoretical one.  It is worth noting that
configuring Juniper VMX for Invisible MPLS tunnels is identical than with Olive.
Invisible tunnels are, now, revealed through DPR, with the \rtla trigger.

\begin{cisco}[title=JunOS VMX -- Invisible PHP]
PE1
no-propagate-ttl

PE2
no-propagate-ttl

P1
no-propagate-ttl

P2
no-propagate-ttl

P3
no-propagate-ttl
\end{cisco}

\begin{cisco}[title=\tnt running over JunOS VMX -- Invisible PHP]
Launching TNT: 192.168.2.102 (192.168.2.102)

  1  CE1 ( 172.16.0.5)  <255,64> [frpla = 0][qttl = 1][uturn = 0] 0.96 ms
  2  PE1 ( 172.16.0.2)  <254,63> [frpla = 0][qttl = 1][uturn = 0] 1.66 ms

   RTLA | Length estimation : 3 | Revealed : 3 (difference : 0)
     2.1 [REVEALED] left.P1 (192.168.1.2)  <253,62> [frpla = 0][qttl = 1][uturn = 0] 8.8 ms  - step 0
     2.2 [REVEALED] left.P2 (192.168.1.6)  <252,62> [frpla = 0][qttl = 1][uturn = 0] 2.134 ms  - step 0
     2.3 [REVEALED] left.P3 (192.168.1.10)  <251,62> [frpla = 0][qttl = 1][uturn = 0] 3.352 ms  - step 0

  3  left.PE2 (192.168.1.14)  <250,62> [frpla = 3][rtl = 3(3)][qttl = 1][uturn = 3] 4.569 ms
  4  CE2 (192.168.2.2)  <250,61> [frpla = 2][rtl = 2(-1)][qttl = 1][uturn = 2] 4.625 ms
  5  CE3 (192.168.2.102)  <250,60> [frpla = 1][rtl = 1(-1)][qttl = 1][uturn = 1] 4.355 ms
\end{cisco}


\subsection{Corner Cases: Heterogeneous Propagation Configuration}\label{appendix.corner}
This section discusses corner cases, i.e., unlikely configurations that may
arise when MPLS is not homogeneously configured throughout the tunnel.  \tnt,
like \traceroute, cannot deal with those situations, but these abnormal shiftings
have not been clearly encountered in practice.

\subsubsection{Cisco Jumpy Configurations}\label{appendix.corner.cisco}
The following Cisco configuration (for IOS 15.2) is supposed to build an UHP
Invisible tunnel.  However, on the contrary to the configuration provided in
Appendix~\ref{appendix.invisible.cisco}, the management of LSE-TTL is
heterogeneous over the tunnel.  Indeed, in this case, the Ingress LER is not
configured with the \notpropagate (on the contrary to the Egress LER and other
routers in the tunnel).  As such, the \minttl{IP-TTL}{LSE-TTL} operation is not
-- systematically -- applied on the Egress while it is expected to be from the
Ingress. The EH assumes that the propagation configuration is homogeneous among
LERs, which is not the case here.  Therefore, the Egress LER will use the IP-TTL
instead of the LSE-TTL when popping the LSE.  As consequence, and as shown by
the \tnt output, we observe that
\begin{enumerate}
  \item the MPLS tunnel is actually Explicit;
  \item a number of hops equal to the tunnel length after the MPLS
  tunnel are missing (here, only CE2 is missing as the platform is too short -- see
  Fig.~\ref{appendix.topo.cisco} for the Cisco topology we use), leading to a
  so-called \dfn{jump} effect.
\end{enumerate}

We call such a configuration \dfn{Explicit Jump} and it can be observed in the
qTTL of the last hop (2 instead of one plus the skipped hop).

\begin{cisco}[title=IOS 15.2 -- Explicit Jump (heterogeneous configuration)]
PE1
version 15.2
mpls label protocol ldp
mpls ldp explicit-null
router bgp 3333
 redistribute connected
 redistribute ospf 10
 neighbor 10.12.0.1 remote-as 3333
 neighbor 10.12.0.1 next-hop-self
 neighbor 192.168.8.1 remote-as 1024
 neighbor 192.168.8.1 next-hop-self

PE2
version 15.2
mpls label protocol ldp
no  propagate-ttl
mpls ldp explicit-null
router bgp 3333
 redistribute connected
 redistribute ospf 10
 neighbor 10.12.0.1 remote-as 3333
 neighbor 10.12.0.1 next-hop-self
 neighbor 192.168.2.2 remote-as 2048
 neighbor 192.168.2.2 next-hop-self

P1
version 15.2
mpls label protocol ldp
no  propagate-ttl
mpls ldp explicit-null
router bgp 3333
 neighbor 10.12.0.1 remote-as 3333

P2
version 15.2
mpls label protocol ldp
no  propagate-ttl
mpls ldp explicit-null
router bgp 3333
 neighbor 10.12.0.1 remote-as 3333

P3
version 15.2
mpls label protocol ldp
no  propagate-ttl
mpls ldp explicit-null
router bgp 3333
 neighbor 10.12.0.1 remote-as 3333
\end{cisco}

\begin{cisco}[title=\tnt running over IOS 15.2 -- Explicit Jump (heterogeneous configuration)]
Launching TNT: 192.168.7.1 (192.168.7.1)

  1  left.CE1 (192.168.3.2)  <255,255> [frpla = 0][qttl = 1][uturn = 0] 8.407 ms
  2  left.PE1 (192.168.8.2)  <254,254> [frpla = 0][qttl = 1][uturn = 0] 29.477 ms
  3  left.P1 (10.1.0.2)  <250,253> [frpla = 3][qttl = 1][uturn = 3][MPLS LSE | Label : 19 | LSE-TTL : 1]  79.929 ms
  4  left.P2 (10.2.0.2)  <250,252> [frpla = 2][qttl = 2][uturn = 2][MPLS LSE | Label : 20 | LSE-TTL : 1]  80.573 ms
  5  left.P3 (10.3.0.2)  <250,251> [frpla = 1][qttl = 3][uturn = 1][MPLS LSE | Label : 20 | LSE-TTL : 1]  109.577 ms
  6  left.PE2 (10.4.0.2)  <250,250> [frpla = 0][qttl = 1][uturn = 0] 79.766 ms
  7  192.168.4.2 (192.168.4.2)  <250,250> [frpla = -1][qttl = 2][uturn = 0] 109.357 ms
\end{cisco}

\subsubsection{Juniper Jumpy Configurations}\label{appendix.corner.juniper}
In the fashion of Cisco, Juniper with the Olive OS (this is not possible with
VMX) allows to configure an Explicit Jump tunnel with PHP.  The configuration provided
below shows such an MPLS tunnel. The EH is configured with the
\notpropagate option, while other routers are configured with \tpropagate.  As
such, P3 will not apply the \minttl{IP-TTL}{LSE-TTL} when popping the label,
leading so to a jump effect that is nearly as long as the tunnel itself (the
Egress LER and CE2 are missing plus the qTTl at 2 on the last hop).

\begin{cisco}[title=Olive -- Explicit Jump (heterogeneous configuration)]
PE1
propagate ttl

PE2
propagate ttl

P1
propagate ttl

P2
propagate-ttl

P3
no-propagate-ttl
\end{cisco}

\begin{cisco}[title=\tnt running over Olive -- Explicit (heterogeneous configuration)]
Launching TNT: 192.168.2.102 (192.168.2.102)

  1  CE1 (172.16.0.5) <255,64> [frpla = 0][qttl = 1][uturn = 0] 0.622 ms
  2  PE1 (172.16.0.2) <254,63> [frpla = 0][qttl = 1][uturn = 0] 1.749 ms
  3  left.P1 (192.168.1.2) <253,62> [frpla = 0][qttl = 1][uturn = 0][MPLS LSE | Label : 299824 | LSE-TTL : 1] 2.799 ms
  4  left.P2 (192.168.1.6) <252,252> [frpla = 0][qttl = 1][uturn = 0][MPLS LSE | Label : 299792 | LSE-TTL : 1] 3.725 ms
  5  left.P3 (192.168.1.10) <251,251> [frpla = 0][qttl = 1][uturn = 0][MPLS LSE | Label : 299776 | LSE-TTL : 1] 7.784 ms
  6  CE3 (192.168.2.102)  <248,57> [frpla = 2][qttl = 2][uturn = 0] 8.884 ms
\end{cisco}

The last configuration is Juniper Olive with an \dfn{Invisible Jump}
configuration.  This is somewhat equivalent to the Explicit Jump but for
Invisible tunnels.  In that case, when P3 (PHP is configured) will pop the LSE,
it will not apply the \minttl{IP-TTL}{LSE-TTL}.  As a result, \tnt will see the
Ingress LER (PE1) and several hops after P3 will be missed (Egress LER and CE2).
The tunnel is invisible and triggers do not work. One can notice a qTTL of 250
on the last hop of our platform: it means that \traceroute can miss an entire
path of 255 minus the length of the tunnel!

\begin{cisco}[title=Olive -- Invisible Jump configuration (heterogeneous configuration)]
PE1
no-propagate ttl

PE2
propagate ttl

P1
no-propagate ttl

P2
no-propagate-ttl

P3
propagate-ttl
\end{cisco}

\begin{cisco}[title=\tnt running over Olive -- Invisible Jump (heterogeneous configuration)]
Launching TNT: 192.168.2.102 (192.168.2.102)
  1  CE1 ( 172.16.0.5)  <255,64> [frpla = 0][qttl = 1][uturn = 0] 0.515 ms
  2  PE1 ( 172.16.0.2)  <254,63> [frpla = 0][qttl = 1][uturn = 0] 1.712 ms
  3  CE3 (192.168.2.102)  <251,60> [frpla = 2][qttl = 250][uturn = 0] 8.553 ms
\end{cisco}

\subsection{RSVP-TE (All Invisible)}\label{appendix.rsvpte}
In this section, we introduce PHP and UHP invisible configurations with both
Cisco and Juniper OSes but considering RSVP-TE instead of LDP.
There is no much to say as all the resulting behaviors are the same as with LDP
configurations described previously. The conclusion is then simple: \tnt is able
to reveal TE tunnels as LDP ones whatever the OS and the popping configuration
is (i.e. PHP or UHP).

\subsubsection{Cisco Config -- PHP}\label{appendix.rsvptephp.cisco}
\begin{cisco}[title=cisco IOS 15.2 -- RSVP-TE PHP]
PE1
version 15.2
no propagate-ttl
mpls traffic-eng tunnels

interface Tunnel0
 ip unnumbered Loopback0
 tunnel mode mpls traffic-eng
 tunnel destination 10.9.0.1
 tunnel mpls traffic-eng autoroute announce
 tunnel mpls traffic-eng priority 1 1
 tunnel mpls traffic-eng bandwidth 500
 tunnel mpls traffic-eng path-option 1 dynamic

PE2
version 15.2
no propagate-ttl
mpls traffic-eng tunnels

P1
version 15.2
no propagate-ttl
mpls traffic-eng tunnels

P2
version 15.2
no propagate-ttl
mpls traffic-eng tunnels

P3
version 15.2
no propagate-ttl
mpls traffic-eng tunnels
\end{cisco}

\begin{cisco}[title=\tnt running over 15.2 -- RSVP-TE PHP]
Launching TraceTunnel: 192.168.7.1 (192.168.7.1)

1  192.168.3.2 (192.168.3.2)  <255,255> [ frpla = 0 ][ qttl = 1 ][ uturn = 0 ]
2  192.168.8.2 (192.168.8.2)  <254,254> [ frpla = 0 ][ qttl = 1 ][ uturn = 0 ]

  FRPLA | Length estimation : 3 | Revealed : 3 (difference : 0)
   2.1 [REVEALED] 10.1.0.2 (   10.1.0.2)  <253,253> [ frpla = 0 ][ qttl = 1 ][ uturn = 0 ]  - step 2
   2.2 [REVEALED] 10.2.0.2 (   10.2.0.2)  <252,252> [ frpla = 0 ][ qttl = 1 ][ uturn = 0 ]  - step 1
   2.3 [REVEALED] 10.3.0.2 (   10.3.0.2)  <251,251> [ frpla = 0 ][ qttl = 1 ][ uturn = 0 ]  - step 0

3  10.4.0.2 (   10.4.0.2)  <250,250> [ frpla = 3 ][ qttl = 1 ][ uturn = 0 ]
4  CE2 (192.168.2.2)  <250,250> [ frpla = 2 ][ qttl = 1 ][ uturn = 0 ]
5  192.168.4.2 (192.168.4.2)  <250,250> [ frpla = 1 ][ qttl = 1 ][ uturn = 0 ]
\end{cisco}

The tunnel is easily revealed as for LDP tunnels.

\subsubsection{Juniper Config -- PHP}\label{appendix.rsvptephp.juniper}

\begin{cisco}[title=Juniper -- RSVP-TE PHP]
PE1
protocols {
    rsvp {
        tunnel-services;
        interface all;
    }
    mpls {
        no-propagate-ttl;
        label-switched-path PE1-to-PE2 {
            to 192.168.1.105;
        }
        interface all;
    }
    ospf {
       traffic-engineering {
           shortcuts lsp-metric-into-summary;
    	}
    }
}

PE2
protocols {
    rsvp {
        tunnel-services;
        interface all;
    }
    mpls {
        no-propagate-ttl;
        icmp-tunneling;
        label-switched-path PE2-toPE1 {
            to 192.168.1.101;
        }
        interface all;
        interface ge-0/0/2.0;
    }
    ospf {
        traffic-engineering;
    }
}

P1
protocols {
    rsvp {
        interface all;
    }
    mpls {
        no-propagate-ttl;
        interface ge-0/0/1.0;
        interface ge-0/0/3.0;
    }
    ospf {
        traffic-engineering;
    }
}

P2
protocols {
    rsvp {
        interface all;
    }
    mpls {
        no-propagate-ttl;
        interface ge-0/0/1.0;
        interface ge-0/0/3.0;
    }
    ospf {
        traffic-engineering;
    }
}

P3
protocols {
    rsvp {
        interface all;
    }
    mpls {
        no-propagate-ttl;
        interface ge-0/0/1.0;
        interface ge-0/0/3.0;
    }
    ospf {
        traffic-engineering;
    }
}
\end{cisco}

\begin{cisco}[title=\tnt running over Juniper -- RSVP-TE PHP]
Launching TraceTunnel: 192.168.2.102 (192.168.2.102)
1  CE1 ( 172.16.0.5)  <255,64> [ frpla = 0 ][ rtla = 0(0) ][ qttl = 1 ][ uturn = 0 ]
2  PE1 ( 172.16.0.2)  <254,63> [ frpla = 0 ][ rtla = 0(0) ][ qttl = 1 ][ uturn = 0 ]

  RTL | Length estimation : 3 | Revealed : 3 (difference : 0)
   2.1 [REVEALED] left.P1 (192.168.1.2)  <253,62> [ frpla = 0 ][ rtla = 0(0) ][ qttl = 1 ][ uturn = 0 ] - step 0
   2.2 [REVEALED] left.P2 (192.168.1.6)  <252,61> [ frpla = 0 ][ rtla = 0(0) ][ qttl = 1 ][ uturn = 0 ] - step 0
   2.3 [REVEALED] left.P3 (192.168.1.10)  <251,60> [ frpla = 0 ][ rtla = 0(0) ][ qttl = 1 ][ uturn = 0 ] - step 0

3  left.PE2 (192.168.1.14)  <250,62> [ frpla = 3 ][ rtla = 3(3) ][ qttl = 1 ][ uturn = 3 ]
4  CE2 (192.168.2.2)  <250,61> [ frpla = 2 ][ rtla = 0(-1) ][ qttl = 1 ][ uturn = 0 ]
5  CE3 (192.168.2.102)  <250,60> [ frpla = 1 ][ rtla = 0(0) ][ qttl = 1 ][ uturn = 0 ]

\end{cisco}

\subsubsection{Cisco Config -- UHP}\label{appendix.rsvpteuhp.cisco}
\begin{cisco}[title=cisco IOS 15.2 -- RSVP-TE UHP]
PE1
version 15.2
no propagate-ttl
mpls traffic-eng tunnels

interface Tunnel0
 ip unnumbered Loopback0
 tunnel mode mpls traffic-eng
 tunnel destination 10.9.0.1
 tunnel mpls traffic-eng autoroute announce
 tunnel mpls traffic-eng priority 1 1
 tunnel mpls traffic-eng bandwidth 500
 tunnel mpls traffic-eng path-option 1 dynamic

PE2
version 15.2
no propagate-ttl
mpls traffic-eng tunnels
mpls ldp explicit-null

P1
version 15.2
no propagate-ttl
mpls traffic-eng tunnels

P2
version 15.2
no propagate-ttl
mpls traffic-eng tunnels

P3
version 15.2
no propagate-ttl
mpls traffic-eng tunnels
mpls traffic-eng signalling interpret explicit-null verbatim
\end{cisco}

\begin{cisco}[title=\tnt running over 15.2 -- RSVP-TE UHP]
Launching TraceTunnel: 192.168.7.1 (192.168.7.1)
1  192.168.3.2 (192.168.3.2)  <255,255> [ frpla = 0 ][ qttl = 1 ][ uturn = 0 ]
2  192.168.8.2 (192.168.8.2)  <254,254> [ frpla = 0 ][ qttl = 1 ][ uturn = 0 ]

  Duplicate IP (Egress : 10.4.0.2) | Length estimation : 3 | Revealed : 4 (difference : 1)
   2.1 [REVEALED] 10.1.0.2 (   10.1.0.2)  <253,253> [ frpla = 0 ][ qttl = 1 ][ uturn = 0 ] - step 4
   2.2 [REVEALED] 10.2.0.2 (   10.2.0.2)  <252,252> [ frpla = 0 ][ qttl = 1 ][ uturn = 0 ] - step 3
   2.3 [REVEALED] 10.3.0.2 (   10.3.0.2)  <251,251> [ frpla = 0 ][ qttl = 1 ][ uturn = 0 ] - step 2
   2.4 [REVEALED] 10.4.0.2 (   10.4.0.2)  <250,250> [ frpla = 0 ][ qttl = 1 ][ uturn = 0 ] - step 1  ( Buddy used )

3  CE2 (192.168.2.2)  <250,250> [ frpla = 3 ][ qttl = 1 ][ uturn = 0 ][ meta = 3, 0, 0 ]
4  CE2 (192.168.2.2)  <250,250> [ frpla = 2 ][ qttl = 1 ][ uturn = 0 ][ meta = -1, 0, 0 ]
5  192.168.4.2 (192.168.4.2)  <250,250> [ frpla = 1 ][ qttl = 1 ][ uturn = 0 ][ meta = 0, 0, 0 ]
\end{cisco}

\subsubsection{Juniper Config -- UHP}\label{appendix.rsvpteuhp.juniper}
\begin{cisco}[title=Juniper -- RSVP-TE UHP]
PE1
protocols {
    rsvp {
        tunnel-services;
        interface all;
    }
    mpls {
        no-propagate-ttl;
        label-switched-path PE1-to-PE2 {
            to 192.168.1.105;
        }
        interface all;
    }
    ospf {
       traffic-engineering {
           shortcuts lsp-metric-into-summary;
       }
    }
}

PE2
protocols {
    rsvp {
        tunnel-services;
        interface all;
    }
    mpls {
        no-propagate-ttl;
        explicit-null;
        icmp-tunneling;
        label-switched-path PE2-toPE1 {
            to 192.168.1.101;
        }
        interface all;
        interface ge-0/0/2.0;
    }
    ospf {
        traffic-engineering;
    }
}

P1
protocols {
    rsvp {
        interface all;
    }
    mpls {
        no-propagate-ttl;
        interface ge-0/0/1.0;
        interface ge-0/0/3.0;
    }
    ospf {
        traffic-engineering;
    }
}

P2
protocols {
    rsvp {
        interface all;
    }
    mpls {
        no-propagate-ttl;
        interface ge-0/0/1.0;
        interface ge-0/0/3.0;
    }
    ospf {
        traffic-engineering;
    }
}

P3
protocols {
    rsvp {
        interface all;
    }
    mpls {
        no-propagate-ttl;
        interface ge-0/0/1.0;
        interface ge-0/0/3.0;
    }
    ospf {
        traffic-engineering;
    }
}
\end{cisco}

\begin{cisco}[title=\tnt running over Juniper -- RSVP-TE UHP]
Launching TraceTunnel: 192.168.2.102 (192.168.2.102)
1  CE1 ( 172.16.0.5)  <255,64> [ frpla = 0 ][ rtla = 0(0) ][ qttl = 1 ][ uturn = 0 ]
2  PE1 ( 172.16.0.2)  <254,63> [ frpla = 0 ][ rtla = 0(0) ][ qttl = 1 ][ uturn = 0 ]

  RTL | Length estimation : 3 | Revealed : 3 (difference : 0)
   2.1 [REVEALED] left.P1 (192.168.1.2)  <253,62> [ frpla = 0 ][ rtla = 0(0) ][ qttl = 1 ][ uturn = 0 ] - step 0
   2.2 [REVEALED] left.P2 (192.168.1.6)  <252,61> [ frpla = 0 ][ rtla = 0(0) ][ qttl = 1 ][ uturn = 0 ] - step 0
   2.3 [REVEALED] left.P3 (192.168.1.10)  <251,60> [ frpla = 0 ][ rtla = 0(0) ][ qttl = 1 ][ uturn = 0 ] - step 0

3  left.PE2 (192.168.1.14)  <250,62> [ frpla = 3 ][ rtla = 3(3) ][ qttl = 1 ][ uturn = 3 ]
4  CE2 (192.168.2.2)  <250,61> [ frpla = 2 ][ rtla = 0(-1) ][ qttl = 1 ][ uturn = 0 ]]
5  CE3 (192.168.2.102)  <250,60> [ frpla = 1 ][ rtla = 0(0) ][ qttl = 1 ][ uturn = 0 ]
\end{cisco}

Again, as one can observe, there is no difference in the trace revelation
regarding the LDP case (in any configurations).

\section{P2MP Circuits (All Invisible)}\label{appendix.p2mp}
In this section, we study the VPRN case. In addition to MPLS, it requires the
MP-BGP feature to build P2MP tunnels. We divide the analysis in two invisible
tunnels sub-cases: the Implicit Null model (relying on the PHP option) and the
Explicit Null one (enabling the UHP option). Note that we simplify a bit the
given configurations for obvious readability purposes (as in the previous TE
section). Moreover the IP configuration is not exactly the same as for previous
cases as its is a more complex MPLS usage.

\subsection{VPN BGP MPLS -- Implicit Null Model}\label{appendix.vpnphp}

\subsubsection{Cisco Config -- PHP}\label{appendix.vpnphp.cisco}
\begin{cisco}[title=cisco IOS 15.2 -- MPLS BGP VPN PHP]
PE1
version 15.2
no propagate-ttl
mpls label protocol ldp

ip vrf VPN_Y
 rd 1:2
 route-target export 1:2
 route-target import 1:2

interface GigabitEthernet4/0
 description PEtoCEVPN
 ip vrf forwarding VPN_Y

router rip
  address-family ipv4 vrf VPN_Y
  redistribute bgp 1 metric 1
  network 10.0.0.0

router bgp 1
 address-family vpnv4
 neighbor 10.0.0.131 activate
 neighbor 10.0.0.131 send-community extended
 neighbor 10.0.0.132 activate
 neighbor 10.0.0.132 send-community extended
 neighbor 10.0.0.130 activate
 neighbor 10.0.0.130 send-community extended

 address-family ipv4 vrf VPN_Y
 redistribute rip
 no synchronization

PE2
version 15.2
no propagate-ttl
ip vrf VPN_Y
 rd 1:2
 route-target export 1:2
 route-target import 1:2

interface GigabitEthernet4/0
 description PEtoCEVPN
 ip vrf forwarding VPN_Y

router rip
 address-family ipv4 vrf VPN_Y
 redistribute bgp 1 metric 1
 network 10.0.0.0

router bgp 1
 address-family vpnv4
 neighbor 10.0.0.130 activate
 neighbor 10.0.0.130 send-community extended
 neighbor 10.0.0.132 activate
 neighbor 10.0.0.132 send-community extended
 neighbor 10.0.0.133 activate
 neighbor 10.0.0.133 send-community extended

 address-family ipv4 vrf VPN_Y
 redistribute rip
 no synchronization
\end{cisco}

\begin{cisco}[title=\tnt running over 15.2 -- VPN BGP MPLS PHP]
  Launching TraceTunnel: 10.0.1.103 (10.0.1.103)
    1  192.168.3.2 (192.168.3.2)  <255,255> [ frpla = 0 ][ qttl = 1 ][ uturn = 0 ]
    2  10.0.0.54 (  10.0.0.54)  <254,254> [ frpla = 0 ][ qttl = 1 ][ uturn = 0 ]

      OPAQUE | Length estimation : 3 | Revealed : 0 (difference : 3)

    3  10.0.0.58 (  10.0.0.58)  <250,250> [ frpla = 3 ][ qttl = 1 ][ uturn = 0 ][MPLS LSE | Label : 28 | mTTL : 252 ]
    4  10.0.0.57 (  10.0.0.57)  <249,249> [ frpla = 3 ][ qttl = 1 ][ uturn = 0 ]
\end{cisco}

The bottom label is revealed and equal to 252. This is the only visible MPLS
indication. No revelation is working on such Opaque tunnels.
Whatever the kind of probes sent to or through the VPRN, the IP address visible
to \tnt (or \texttt{traceroute} in general) is the outgoing address. Despite its
expired TTL, it is likely that the probe arriving on the Egress PE will be
pushed to the VRF of the VPN and its associated interface before generating the
error message (the VPN being identified with the MPLS label contained in the
packet). Then, the interface where the packet actually expires is the one
associated to the VRF. However, we are able to distinguish UHP and PHP
configurations (thanks to the so called LSE-TTL++), because the bottom label is
equal to $255$ for UHP and lower with PHP as we can see here. Note that the two
last IP addresses can also trigger an alarm. In any cases, we are able to
discriminate them from other opaque tunnels shown in previous P2P
configurations. So we are able to identify their class as they are not possible
to reveal.

\subsubsection{Juniper Config -- PHP}\label{appendix.vpnphp.juniper}
\begin{cisco}[title=Juniper -- VPN BGP MPLS PHP]
PE1
routing-instances
    VRF1
        instance-type vrf;
        interface ge-0/0/2.0;
        route-distinguisher 192.168.1.101:1;
        vrf-target target:65000:1;
        protocols
            bgp
                group ce
                    type external;
                    peer-as 1;
                    neighbor 172.16.0.1;
protocols
  mpls
    interface all;
    no-propagate-ttl;

    bgp
        group internal-peers
            type internal;
            local-address 192.168.1.101;
            family inet-vpn
                any;
            export next-hop-self;
            neighbor 192.168.1.106;
            neighbor 192.168.1.105;

PE2
routing-instances
    VRF1
        instance-type vrf;
        interface ge-0/0/2.0;
        route-distinguisher 192.168.1.105:1;
        vrf-target target:65000:1;
        protocols
            bgp
                group ce
                    type external;
                    peer-as 3;
                    neighbor 192.168.2.2;
protocols
    mpls
        no-propagate-ttl;
        icmp-tunneling;
        interface all;

    bgp
        group internal-peers
            type internal;
            local-address 192.168.1.105;
            family inet-vpn
                any;
            export next-hop-self;
            neighbor 192.168.1.106;
            neighbor 192.168.1.101;

\end{cisco}

\begin{cisco}[title=\tnt running over Juniper towards the buddy -- BGP MPLS VPN PHP]
Launching TraceTunnel: 192.168.2.102 (192.168.2.102)
1  CE1 ( 172.16.0.5)  <255,64> [ frpla = 0 ][ rtla = 0(0) ][ qttl = 1 ][ uturn = 0 ]
2  PE1 ( 172.16.0.2)  <254,63> [ frpla = 0 ][ rtla = 0(0) ][ qttl = 1 ][ uturn = 0 ]

  RTL | Length estimation : 3 | Revealed : 1 (difference : 2)
   2.1 CE2 (192.168.2.2)  <250,62> [ frpla = 0 ][ rtla = 0(0) ][ qttl = 1 ][ uturn = 0 ] - step 1  ( Buddy used )

3  CE2 (192.168.2.2)  <250,62> [ frpla = 3 ][ rtla = 3(3) ][ qttl = 1 ][ uturn = 3 ]
4  CE3 (192.168.2.102)  <250,61> [ frpla = 2 ][ rtla = 0(0) ][ qttl = 1 ][ uturn = 0 ]
\end{cisco}

We can observe that CE2 appears twice. When targeting the buddy (192.168.2.1),
we \textit{rediscover} CE2 since it now appears as if it was before the buddy in
the topology.  We say that we capture a \textit{twisted IP} when targeting the
buddy.

\begin{cisco}[title=\tnt running over Juniper -- BGP MPLS VPN PHP]
Launching TraceTunnel: 192.168.2.1 (192.168.2.1)
1  CE1 ( 172.16.0.5)  <255,64> [ frpla = 0 ][ rtla = 0(0) ][ qttl = 1 ][ uturn = 0 ]
2  PE1 ( 172.16.0.2)  <254,63> [ frpla = 0 ][ rtla = 0(0) ][ qttl = 1 ][ uturn = 0 ]

  RTL | Length estimation : 3 | Revealed : 1 (difference : 2)
   2.1 CE2 (192.168.2.2)  <250,62> [ frpla = 0 ][ rtla = 0(0) ][ qttl = 1 ][ uturn = 0 ] - step 1  ( Buddy used )

3  CE2 (192.168.2.2)  <250,62> [ frpla = 3 ][ rtla = 3(3) ][ qttl = 1 ][ uturn = 3 ]
4  192.168.2.1 (192.168.2.1)  <250,63> [ frpla = 2 ][ rtla = 4(0) ][ qttl = 1 ][ uturn = 4 ]
\end{cisco}

With Juniper VPN, there is no Opaque indicator resulting from VPRN or any other
configurations. A first explanation is that Juniper routers, on the contrary to
the independent mode enabled by default with Cisco routers, do not inject the
whole IGP in LDP, but only their loopback address using the ordered mode (see
Sec.~\ref{background}). This mode limits the probability to face a
non-controlled tunnel ending. However, with VPRN configurations, a Juniper
Egress LER deals with the same packet level situation as with Cisco routers. Up
to the end of the tunnel, the packet is still MPLS encapsulated with the
end-to-end VPN non terminating label at the bottom of the stack. Juniper routers
do not, however, produce an Opaque indicator in that situation. Indeed, packets
destined to the VPN are handled in a specific way with Juniper devices: they are
IP packets forwarded directly to the next-hop without looking at or manipulating
the IP-TTL whatever its value.

The outcome of such a sliding packet is twofold. Firstly, the Egress hop is
hidden in the transit trace, as with Cisco UHP but without the duplicated IP.
Secondly, when performing a direct trace (even with UDP) targeting the first
address of the path within the VPN, i.e the IP interface of the Egress LER
belonging to the VPN, one can see that this address and its buddy appear in the
wrong order. Indeed, in the trace, the two addresses are switched, meaning that
the CE IP address appears before the Egress one. Being forwarded without
inspecting the IP-TTL, probes targeting IP addresses belonging to the VPN are
automatically forwarded to the CE router, where they expire. The next probe,
having a greater TTL, follows the same path as the one before, but can be
forwarded back to the Egress LER by the CE router before expiring. This loop
results in the two addresses being switched regarding their actual location in
the path. Finally, one can infer the loop because two additional artifacts
compared to RTLA (RTLA++) are visible: the TTL that deviates from its monotony and
subsequent IP addresses also raise alarms due to potential conflicting
allocation.

\subsection{VPN BGP MPLS -- Explicit Null}\label{appendix.vpnuhp}
\subsubsection{Cisco Config -- UHP}\label{appendix.vpnuhp.cisco}

\begin{cisco}[title=cisco IOS 15.2 -- MPLS BGP VPN PHP]
PE1
version 15.2
no propagate-ttl
mpls label protocol ldp

ip vrf VPN_Y
 rd 1:2
 route-target export 1:2
 route-target import 1:2

interface GigabitEthernet4/0
 description PEtoCEVPN
 ip vrf forwarding VPN_Y

router rip
  address-family ipv4 vrf VPN_Y
  redistribute bgp 1 metric 1
  network 10.0.0.0

router bgp 1
 address-family vpnv4
 neighbor 10.0.0.131 activate
 neighbor 10.0.0.131 send-community extended
 neighbor 10.0.0.132 activate
 neighbor 10.0.0.132 send-community extended
 neighbor 10.0.0.130 activate
 neighbor 10.0.0.130 send-community extended

 address-family ipv4 vrf VPN_Y
 redistribute rip

PE2
version 15.2
no propagate-ttl
mpls ldp explicit-null

ip vrf VPN_Y
 rd 1:2
 route-target export 1:2
 route-target import 1:2

interface GigabitEthernet4/0
 description PEtoCEVPN
 ip vrf forwarding VPN_Y

router rip
 address-family ipv4 vrf VPN_Y
 redistribute bgp 1 metric 1
 network 10.0.0.0

router bgp 1
 address-family vpnv4
 neighbor 10.0.0.130 activate
 neighbor 10.0.0.130 send-community extended
 neighbor 10.0.0.132 activate
 neighbor 10.0.0.132 send-community extended
 neighbor 10.0.0.133 activate
 neighbor 10.0.0.133 send-community extended

 address-family ipv4 vrf VPN_Y
 redistribute rip
\end{cisco}

\begin{cisco}[title=\tnt running over 15.2 -- VPN BGP MPLS UHP]
Launching TraceTunnel: 10.0.1.103 (10.0.1.103)
    1  192.168.3.2 (192.168.3.2)  <255,255> [ frpla = 0 ][ qttl = 1 ][ uturn = 0 ]
    2  10.0.0.54 (  10.0.0.54)  <254,254> [ frpla = 0 ][ qttl = 1 ][ uturn = 0 ]

      OPAQUE | Length estimation : 3 | Revealed : 0 (difference : 3)

    3  10.0.0.58 (  10.0.0.58)  <250,250> [ frpla = 3 ][ qttl = 1 ][ uturn = 0 ][MPLS LSE | Label : 28 | mTTL : 255 ]
    4  10.0.0.57 (  10.0.0.57)  <249,249> [ frpla = 3 ][ qttl = 1 ][ uturn = 0 ]
\end{cisco}

The bottom label is revealed and equal to 255. We can identify this tunnel as
the VPRN UHP case but not reveal its content as for non VPRN opaque ones.

To conclude, it appears that VPRN content cannot be revealed with \tnt, while
other Opaque tunnels configurations (i.e., routing devices heterogeneity, BGP
edge configuration) can. The mechanism behind the absence of content revelation
can be explained by the IP address collected by \tnt from the source IP field in
the ICMP reply. Usually, the collected address is the one assigned to the
physical incoming interface of the Egress PE. In the VPRN case, the collected IP
is the one assigned to the interface on which the VRF is attached. In practice,
this corresponds to the outgoing interface towards the VPN at the customer's
side. Said otherwise, \tnt collects the outgoing address instead of the incoming
one. Because the incoming address is the only one that enables a successful
revelation, this type of Opaque tunnels cannot be revealed yet.

\end{document}